# Wake asymmetry weakening in viscoelastic fluids: Numerical discovery and mechanism exploration

Sai Peng, Tao Huang,  Taiba Kouser, et al.

View Online

Export Citation

CrossMark

## ARTICLES YOU MAY BE INTERESTED IN

# Wake asymmetry weakening in viscoelastic fluids: Numerical discovery and mechanism exploration

Sai Peng (彭赛)[1,2], Tao Huang (黄涛)[1,2], Taiba Kouser[3], Xiao-ru Zhuang (庄晓如)[1,4], Yong-liang Xiong (熊永亮)[5], and Peng Yu (余鹏)[1,2*]

[1]Department of Mechanics and Aerospace Engineering, Southern University of Science and Technology, Shenzhen, 518055, China

[2] Center for Complex Flows and Soft Matter Research, Southern University of Science and Technology, Shenzhen, 518055, China

[3]Department of Mathematics, Government College University Faisalabad, 38000, Pakistan

[4]School of Mechanical and Electrical Engineering, Shenzhen Polytechnic, Shenzhen, 518055, China

[5]Department of Mechanics, Huazhong University of Science and Technology, Wuhan 430074, China

## Abstract

Viscoelasticity weakens the asymmetry of laminar shedding flow behind a blunt body in a free domain. In the present study, this finding is confirmed by four unsteady viscoelastic flows with asymmetric flow configuration, i.e., flow over an inclined flat plate with various angles of incidence, flow over a rotating circular cylinder, flow over a circular cylinder with asymmetric slip boundary distribution, and flow over an inclined row of eight equally closely spaced circular cylinders (which can be considered as a single large blunt body) through direct numerical simulation combined with the Peterlin approximation of the finitely extensible nonlinear elastic (FENE-P) model. At high Weissenberg number, an arc shape region with high elastic stress, which is similar to shock wave, forms in the frontal area of the blunt body. This region acts as a stationary shield to separate the flow into different regions. Thus, the free stream resembles to pass this shield instead of the original blunt body. As this shield has symmetric feature, the wake flow restores symmetry.

**Keywords:** Viscoelastic flow, shedding flow asymmetry, inclined flat plate, rotating circular cylinder, asymmetric slip distribution, side-by-side cylinders.

## I. Introduction

When a long cylindrical structure is immersed in a cross-flow, Kármán vortices are periodically shed downstream of the structure at sufficiently large Reynolds number ($Re$).[1] These vortex shedding phenomena from blunt cylinders have received great attention since they are associated with numerous cases of flow-induced structural and acoustic vibrations.[2,3] For the simplest case of unconfined viscous flow past a circular cylinder, depending upon $Re$, the flow undergoes several transitions from one flow regime to another. At very low $Re$, since fluid inertia is negligible, fluid parcels are able to adjust the shape of the submerged blunt body and thus closely follow its contours, i.e., the flow remains attached to the surface. As $Re$ is increased gradually ($Re > \sim 5$), fluid inertia increases and the adverse pressure gradient along the surface of the blunt body leads to the appearance of a separation bubble. The flow remains symmetric with respect to the cylinder center line along the cross-flow direction within this low $Re$ range. With a further increment in $Re$, the separation bubble grows in size until the wake becomes asymmetric and unsteady at $Re \approx 48$. Beyond this $Re$, the vortices are shed downstream of the cylinder and the wake becomes periodic in time, albeit the flow is still two-dimensional and laminar. Although the transient flow shows asymmetry, previous numerical simulation and experimental results indicated that the time-averaged flow field around a cylinder still maintains the spatial symmetry.[4,5] Statistically, the time-averaged lift acted on cylinder is about zero.

**\*Corresponding Author:** yup6@sustech.edu.cn

Flow over blunt bodies is also affected by the shape,[6] orientation,[7] number and arrangement,[8] and motion (such as rotation)[9] of the blunt bodies, boundary condition (such as wall slip),[10] as well as the blockage ratio of flow domain,[11] etc. In Newtonian fluid, when the flow configuration (i.e., geometry or boundary condition) is asymmetric, the pressure distribution on the blunt body and the vortex shedding in the wake show up-down asymmetry. For example, shear layer attachment on an inclined flat plate could lead to an effective change of the flow pattern due to the inclination of the inclined flat plate,[7] which may lead to asymmetric flow beyond a certain angle of attack ($\alpha$). With an increase in $\alpha$, the lift coefficient of inclined flat plate gradually increases. This results from the asymmetric pressure distribution on the upper and lower surfaces of the inclined flat plate. The vortex shedding trajectory in the wake deviates from the center line of the flow field. Over a rotating circular cylinder, the cross flow would also generate lateral lift due to asymmetry. As the cylinder rotation drives additional motion of the surrounding fluid, the fluid velocity increases at one side of the cylinder and decreases at the other side. This difference causes the pressure on the upper and lower sides of the cylinder to be inconsistent, resulting in the force perpendicular to the incoming flow direction, which is the so-called Magnus effect.[12] Correspondingly, the wake flow at $Re = 100$ could be divided into four categories, including two types of vortex shedding modes and two types of steady states.[9] Similar to flow over a rotating circular cylinder, for flow over a circular cylinder with asymmetric slip distributions, no-zero average torque and lift may exist due to the asymmetric flow near the cylinder's surface.[10] Meanwhile, an asymmetric vortex shedding trajectory appears downstream of the cylinder.

Flow over two side-by-side circular cylinders is a simple representation of multi-body system in a free stream. At $Re = 100$ (based on cylinder's diameter $D$), the flow behaviors can be classified into four different modes, depending on $L_D/D$ ($L_D$ is the minimum spacing between two cylinders), i.e., the single vortex street mode when $L_D/D < 0.4$, the flip-flopping mode when $0.4 \leq L_D/D < 1.5$, the in-phase-synchronized flow mode when $1.5 \leq L_D/D < 2.0$, and the antiphase-synchronized flow mode when $L_D/D \geq 2.0$.[8] If the distance between cylinders is small ($L_D/D < 0.4$), the two cylinders can be regarded as a single blunt body and the corresponding flow resembles that past a single blunt body. Moreover, a strong repulsive force exists between the cylinders if they are placed closely.[8] The flow behaviors at the upper and lower sides of each cylinder show different features. For example, when the fluid flows past the upper side of the top cylinder, the velocity gradually recovers the incoming flow velocity along the streamwise direction. However, the flow velocity near the lower side of the top cylinder is much smaller, due to the blockage effect of the narrow gap between the two cylinders. Thus, the near surface flow is asymmetric with respect to the horizontal center line of each cylinder, which results in the nonzero time-averaged lift force acted on each cylinder.

Adding dilute concentration of polymers in a fluid could significantly change the flow characteristics. For the internal and external flows in turbulent flow regime, when a small amount of polymer is added into water, the friction factor and the drag coefficient are known to dramatically decrease, which implies many applications ranging from fluid transportation to flow control.[13,14] For example, Xiong et al.[15] proposed a strategy to suppress vortex-induced vibration by introducing a small amount of soluble long-chain polymer in water. Adding soluble polymer into water could also suppress cavitation.[16-19] Dissolving polymer into a working fluid may either inhibit heat dissipation[20-22] or enhance heat transfer.[23-25] Solutions containing polymer additives may exhibit more complex rheological behavior than Newtonian fluids, owing to the non-Newtonian effects such as shear-thinning, anisotropy, and viscoelasticity. Therefore, the underlying mechanisms behind these applications are very complicated. When the geometry and the boundary conditions are symmetric, flow asymmetry is more likely to occur in viscoelastic fluid, compared with pure shear-thinning fluid or Newtonian fluid flow.[26-28] Haward et al.[26] experimentally studied the flow of a dilute polymer solution (low polydispersity sample of tactic polystyrene dissolved in a viscous organic solvent dioctyl phthalate) over a confined cylinder. This fluid is essentially non-shear-thinning over 3 decades in shear rate. They found that at high Weissenberg number (We), a flow asymmetry appeared upstream of the cylinder, due to a high local tensile Weissenberg number. Later, Haward et al.[27] investigated a series of shear-banding viscoelastic wormlike micellar solutions (hydrolyzed polyacrylamide dissolved at different concentrations in deionized water) and found that strong flow asymmetry appears not only in front of the cylinder but also at both sides of the cylinder. The asymmetry was also found to develop from an initially random sideway fluctuation of the highly-stressed downstream birefringent wake when We is beyond a critical value $We_c$. Besides, the asymmetry also widely appears in other flows, such as cross-flow[29,30], planar expansion,[31] etc.

Xiong et al.[32] simulated viscoelastic flow over a hydrofoil with a large $\alpha$ and found that the wake field gradually retains symmetry (associated with a decreasing lift) when increasing We. For viscoelastic flow over two side-by-side circular cylinders, the lift force between the two cylinders with a low $L_D$ become lower when increasing We.[33] In these two numerical simulations, it is observed that the lift force acted on every blunt body is weakened in viscoelastic fluid, comparing with that in Newtonian fluid. This raises questions as to whether and why the flow asymmetry against each blunt body caused by the asymmetric geometry or boundary condition could be weakened in viscoelastic fluid.

To shed light on this matter, we propose four numerical examples with asymmetric flow configuration against each blunt body, i.e., flow over an inclined flat plate with various angles of incidence $\alpha$, flow over a rotating circular cylinder, flow over a circular cylinder with asymmetric slip boundary distribution, and flow over an inclined row of eight equally closely spaced circular cylinders. In particular, flow over eight equally closely spaced circular cylinders is a flow past multibody system, however, which can be regarded as flow past a whole blunt body with geometric asymmetry. Other multi-body flows with closely placed objects, such as flows over two side-by-side circular cylinders[37] and two staggered circular cylinders,[38] can also be classified in this category.

The above four flow examples can be divided into two groups, i.e., geometric asymmetry, boundary condition asymmetry. A rotating circular cylinder and the asymmetric slip boundary condition on the cylinder surface are the representative of asymmetric boundary condition. Flow over an inclined flat plate and flow over an inclined row of eight equally closely spaced circular cylinders are the representative of geometric asymmetry. Other geometric asymmetry examples include a symmetric hydrofoil with a certain angle of attack,[34, 35] an inclined square cylinder,[36] etc. Considering Newtonian fluid in the above flow examples, the lift force acted on each blunt body is non-zero, while the wake field becomes asymmetry. For comparison, we carefully explore the influence of viscoelasticity on their laminar unsteady wake behaviors, particularly the flow asymmetry.

This remaining of the paper is organized as follows. The governing equations and the numerical methods are presented in Sec. II. The results of lift force and flow characteristics of the four flow examples are presented and the corresponding underlying mechanisms on flow asymmetry weakening are discussed in Sec. III. Finally, the main conclusions and the future outlook are provided in Sec. IV.

## II. Mathematical formulation

While a small amount of polymers are added into water, the incompressible Navier-Stokes (N-S) equations are slightly modified and have the following form on a differential fluid element:[39-43]

$$\frac{\partial u_j}{\partial x_j} = 0,$$

(1)

$$\rho \frac{\partial u_i}{\partial t} + \rho u_j \frac{\partial u_i}{\partial x_j} = -\frac{\partial p}{\partial x_i} + \mu_s \frac{\partial^2 u_i}{\partial x_j^2} + \frac{\mu_p}{\lambda} \frac{\partial \tau_{ij}^p}{\partial x_j},$$

(2)

where the summation convention is used, with $j$ being the summation index and $i$ being the component of the vector equation, $x$ is the coordinate, $t$ is time, $u$ is the velocity, $p$ is the pressure, $\rho$ is the density of fluid, $\mu_s$ and $\mu_p$ are the viscosity contribution from the solvent and the polymer, respectively, $\lambda$ is the relaxation time of polymer, $\tau_{ij}^p$ is the additional polymer stress. The last term in Eq. (2) represents the influence of $\tau_{ij}^p$ due to the elasticity of polymers in the flow. The total viscosity of the solution is defined as $\mu = \mu_p + \mu_s$. The polymer viscosity ratio at vanishing shear rate is defined as $\beta = \mu_p/\mu$, which is a measurement of polymer concentration and molecular characteristics. Eq. (2) restores to the original N-S equations when $\beta = 0$. The polymer stress can be modeled by a molecular-based Peterlin approximation of the finitely extensible nonlinear elastic (FENE-P) model, which describes an individual member of polymers in a dilute concentration as a dumbbell connected with a finitely extensible nonlinear elastic spring by the way of a balance of forces acting on each beads. In this model, the polymeric stress $\tau_{ij}^p$ can be determined using kinetic theory.[44,45]

$$\tau_{ij}^p = \frac{c_{ij}}{1 - \frac{c_{kk}}{L^2}} - \delta_{ij},$$

(3)

where $c_{ij}$ represents the polymer conformation tensor, which is defined as the pre-averaged dyadic product of the polymer end-to-end vector, $\delta_{ij}$ is the Kroenecker delta function, and $L$ is the maximum polymer extensibility, which is normalized by the equilibrium length of a linear spring $(K_bT/H)^{1/2}$ with $T$ the absolute temperature, $K_b$ the Boltzmann constant, and $H$ the Hookean spring constant for an entropic spring. The polymer conformation tensor is governed by the following hyperbolic transport equation:

$$\frac{\partial c_{ij}}{\partial t} + u_k \frac{\partial c_{ij}}{\partial x_k} - c_{ik} \frac{\partial u_j}{\partial x_k} - c_{kj} \frac{\partial u_i}{\partial x_k} = -\frac{\tau_{ij}^p}{\lambda}. \tag{4}$$

It is noted that the FENE-P model is employed in this work because of its ability to properly represent the finite extensibility of the polymer. In order to obtain bounded solutions for problems with high $We$ and large strain rate, the finite extensibility is necessary. Those linear spring models such as the Oldroyd-B model cannot be faithfully used in real engineering problems. Furthermore, the FENE-P model has been widely used in previous studies involving viscoelastic flows at high $Re$ and can successfully produce accurate physical results in these problems. In this paper, $\beta$ is set to be very small ($\beta = 0.1$) to minimize the shear-thinning effect and $L$ is set to 100. Under these parameters, the FENE-P model could describe the rheological behavior of a dilute polymer solution more explicitly.

To solve the above equations numerically, a finite volume commercial solver ANSYS FLUENT associated with a self-developed user defined function (UDF) for the elastic stress transport equation, i.e., Eq. (4), is applied. The QUICK scheme and the second order implicit scheme are adopted to discretize the space and temporal domains, respectively. An artificial diffusion term $\kappa \Delta c$ is added to the right hand side of Eq. (4). This implementation has been proved to be effective to ensure numerical stability.[39,40,46-51] Without the artificial diffusion term, Eq. (4) is a hyperbolic equation. Thus, the free stream condition is naturally involved and boundary conditions for the walls of blunt bodies and the outflow are not required. However, the flow variables are stored at the cell center in the present study and the conformation tensor at the boundary must be known when calculating the gradient of the conformation stress in Eq. (2). In this case, the no-flux boundary condition suggested by Alves *et al.*[58] can be adopted for other boundaries except the inlet. It is expected that the artificial diffusion term has very small effect on the solution of Eq. (4). Therefore, the same no-flux boundary condition is adopted in the present study. The flow behaviors at high $We$ obtained by simulation with the artificial diffusion term are consistent with the experimental results, such as instability suppression,[52-54] vortex shedding frequency decrease[52,55] and drag enhancement at low $Re$ and drag reduction at high $Re$.[56] In this study, $\kappa = 0.1\mu$ is adopted in the numerical examples with high $We$ (except for $\kappa = 0.5\mu$ in the asymmetric slip boundary example) by considering both the numerical stability and accuracy of numerical approximation.[39,40] However, for low $We$ cases, no artificial diffusion term is added. Furthermore, an under-relaxation scheme is utilized to avoid the rapid increase of the numerical error. In all our simulation, time-averaged flow field calculation counts more than 20 time periods after the flow reaches the final state. The time-averaged variable is expressed by putting a bar over the corresponding symbol.

## III. Validation for viscoelastic flow simulation

In order to evaluate the accuracy of the present numerical method for the FENE-P fluid, a comparison is made with the analytical solution for the slit flow.[57] The streamwise velocity profile and the first component of non-dimensional polymeric stress profile at $We = 1$ are both plotted in Fig. 1, which shows excellent agreement between the analytical solutions and the present numerical results.

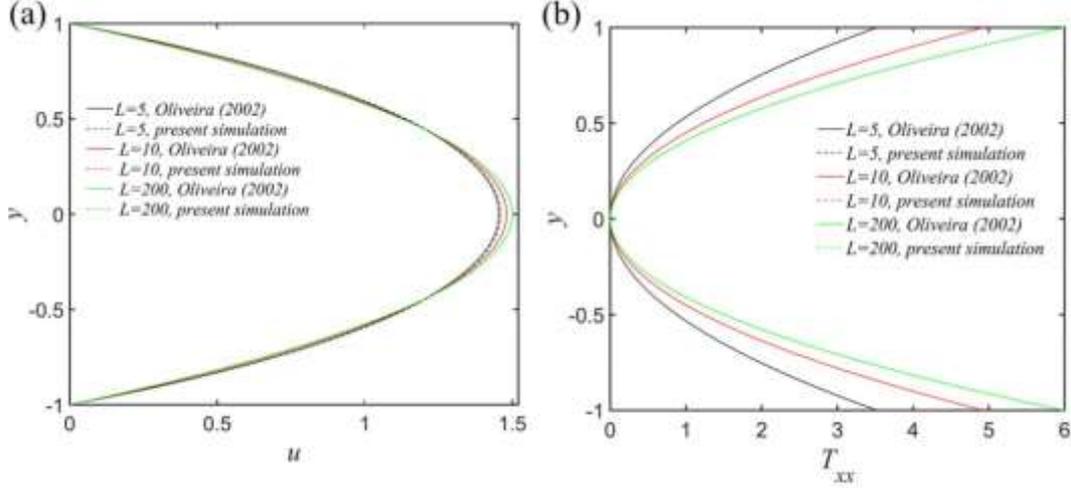

**Fig. 1.** A comparison between the analytical solution and the present simulation results for the slit flow of the FENE-P fluid at $We = 1$, (a) the streamwise velocity $u$ and (b) the first component of non-dimensional polymeric stress $T_{xx}$.

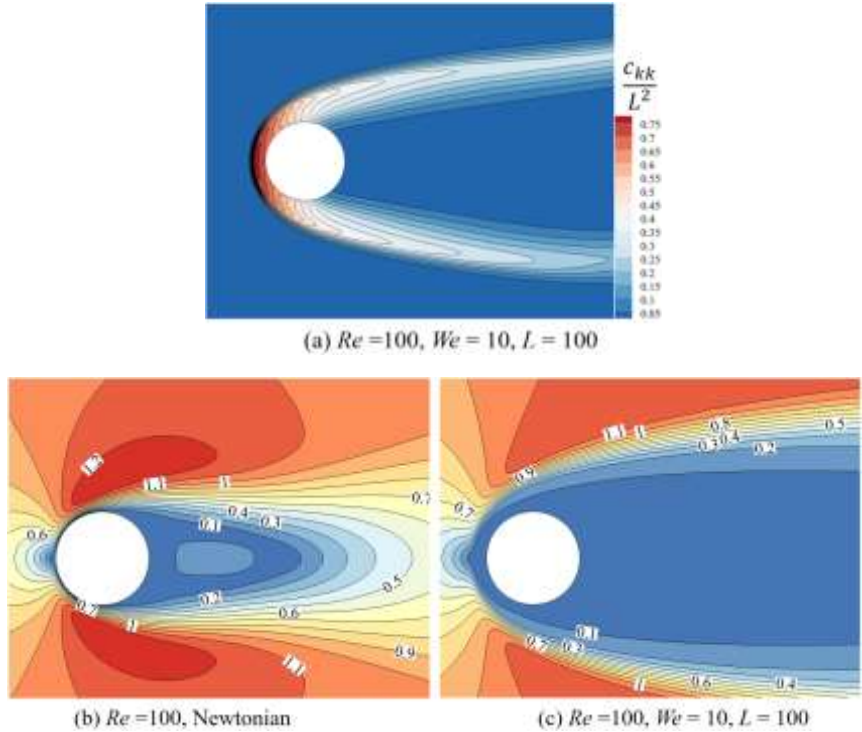

**Fig. 2.** (a) The distribution of the trace of elastic stress tensor ($c_{kk}/L^2$). The distribution of time-averaged flow velocity magnitude ($\sqrt{\bar{u}^2 + \bar{v}^2}/U_\infty$) for (b) Newtonian fluid and (c) viscoelastic fluid with $We = 10$ and $L = 100$.

Furthermore, viscoelastic flow over a circular cylinder at $Re = 100$, $We = 10$ and $L = 100$ in a free domain is simulated. The whole computational domain has a semicircular shape, in which a cylinder with diameter $D$ is embedded. The diameter of the semicircle is $60D$. The center of the cylinder is located at the symmetry line of the semicircle, with the distance of $17D$ to the straight edge. The inlet and the outflow conditions are imposed on the curved edge and the straight edge of the semicircle, respectively. The details of computational domain and boundary conditions could refer to Richter *et al.*[39] Note that this high Weissenberg number problem (HWNP) poses challenges in numerical simulation.[58] In viscoelastic flow at high $We$, viscoelastic stress plays a key role on flow dynamics. In order to understand the relationship between the elastic stress and the velocity (magnitude), the distributions of the trace of elastic stress tensor ($c_{kk}/L^2$) and the time-averaged flow velocity magnitude ($\sqrt{\bar{u}^2 + \bar{v}^2}/U_\infty$) are shown in Fig. 2. The concentrated area of elasticity extends from the upstream to the downstream of the cylinder. The velocity in this concentrated area becomes smaller. The high elastic stress slows down the flow velocity near the cylinder surface.

We compare some details of the flow field and statistical results with those reported in Richter *et al.*[39] Fig. 3 shows the distributions of the elastic stress and the $x$-velocity in front of the cylinder. For high $L$ and high $We$, the flow field shows a solid like behavior in front of the blunt body, characterized by high elastic stress, high pressure, and low flow velocity. The profiles of trace of conformation tensor and the $x$-velocity along the center line demonstrate that our numerical results are consistent with those reported in literature, as shown in Fig. 3. The maximum error for the trace is about 2%, suggesting that the present results only slightly suffer from the numerical diffusion. In addition, it is worth mentioning that the present time-averaged drag coefficient is 2.64, which is very close to that (2.7) reported in Richter *et al.*[39]

In summary, these two validation cases confirm that the capability of the present method to accurately capture the main characteristics of viscoelastic fluid flow considered.

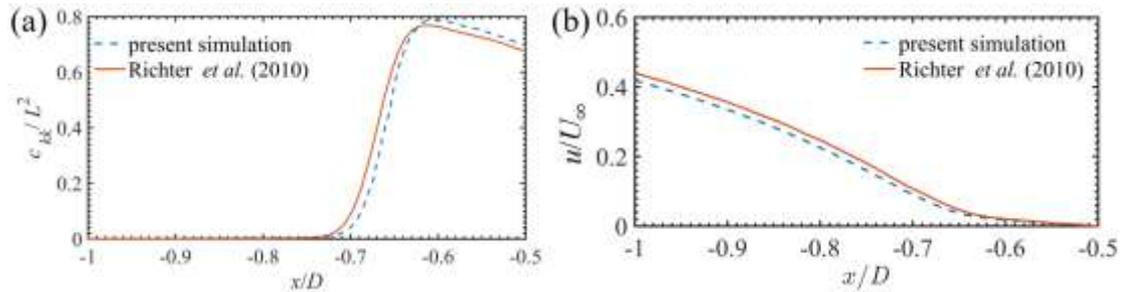

**Fig. 3.** The profiles of (a) the trace of elastic stress tensor ($c_{kk}/L^2$) and (b) the $x$-velocity ($u/U_\infty$) in front of the cylinder along the stagnation streamline. The origin of the coordinate system is set at the forefront of the cylinder.

## IV. Numerical results and discussion

### A. Viscoelastic flow over an inclined flat plate with various angles of incidence

First, viscoelastic flow over an inclined flat plate with various angles of incidence in a free domain is studied, as shown in Fig. 4(a). This special geometry is a representative of geometric asymmetry. The width and the height of the flat plate are set as $a$ and $0.02a$, respectively. The attack incidence between the flat plate and the incoming flow is set as $\alpha$ (10° and 20°). The computational domain is rectangular. The geometrical center of the inclined flat plate is set at the origin of the coordinate system $(x, y) = (0, 0)$. The distance between the inlet and the origin is set as $L_u = 10a$. The distance between the outlet and the origin of coordinate is set as $L_v = 20a$. The distance between the two lateral boundaries is set as $H = 20a$. A uniform streamwise velocity $U_\infty$ and conformation tensor of $\mathbf{c} = \mathbf{I}$ are applied on the inlet boundary. A zero gauge pressure is applied on the outflow boundary with the far field. For the conformation tensor, the no-flux condition is approximated on all boundaries except for the inlet boundary. A slip condition is imposed on the lateral boundaries. Our preliminary test indicates that the difference between the simulation results based on the present computational domain and a double size domain are very small and the influences of the domain size and the lateral boundary condition on the results can be ignored.

The computational grid is generated by the commercial software ICEM. The mesh consists of two parts, which are separated by the red dash lines as shown in Fig. 4. One part includes 8 internal blocks surrounding the inclined flat plate, and the other part includes 16 external blocks extended from the internal blocks. Fig. 4(b) shows the mesh surrounding the flat plate. The width and height of the plate are discretized by 601 and 61 grid points respectively. The mesh resolution of the grid near the four corners of the plate is refined to weaken the influence of concentrated elastic stress. The height of the first cell adjacent to the surface of the plate is set as $0.000625a$ and the grid size increases exponentially with a ratio of 1.02 along the direction normal to the surface to ensure a fine mesh near the plate. For the external blocks in the $x$-direction, 501 (for $L_v$) and 51 (for $L_u$) grid points are unevenly arranged in the downstream region and the upstream region, respectively. For the external blocks in the $y$-direction, the upper and lower regions (separated by the center region) both have 51 unevenly distributed grid points. The whole computational domain includes 271, 800 structured quadrilateral cells in total.

For this numerical example, the Reynolds number is defined as $Re = U_\infty a\rho/\mu$ and fixed at 500 and the Weissenberg number is defined as $We = \lambda U_\infty/a$, which ranges from 0 to 1.5. For most simulation cases, the time step is set to $0.00625a/U_\infty$. For the certain cases that are more difficult to converge (mainly those at high $We$), the time step is reduced to $0.003125a/U_\infty$ or even $0.0015625a/U_\infty$. Our previous studies[15,32,33] indicated that the time step is sufficiently small to obtain reliable and accurate simulation results.

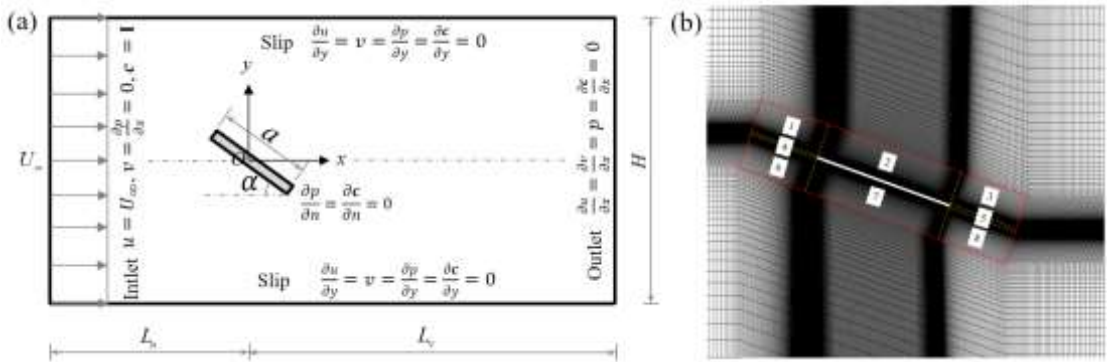

**Fig. 4.** (a) The schematic of the unconfined flow around an inclined flat plate. (b) The mesh topology near the inclined flat plate for $\alpha = 20°$.

Wake asymmetry weakening in viscoelastic fluids

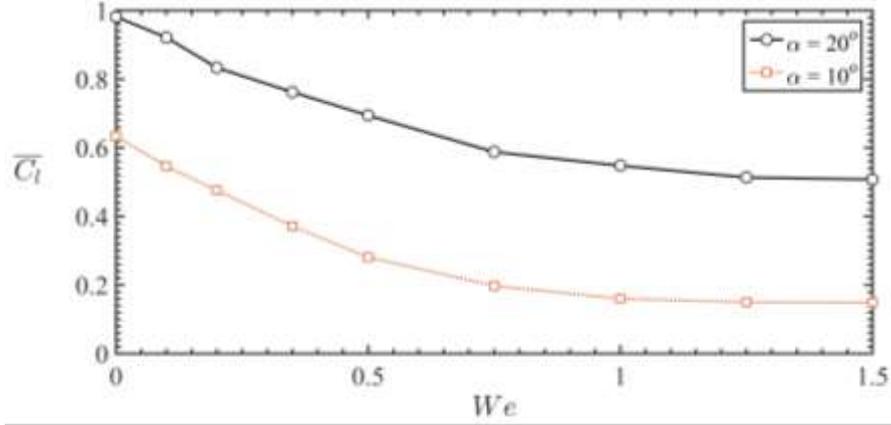

**Fig. 5.** Variation of the time-averaged lift coefficient with *We*.

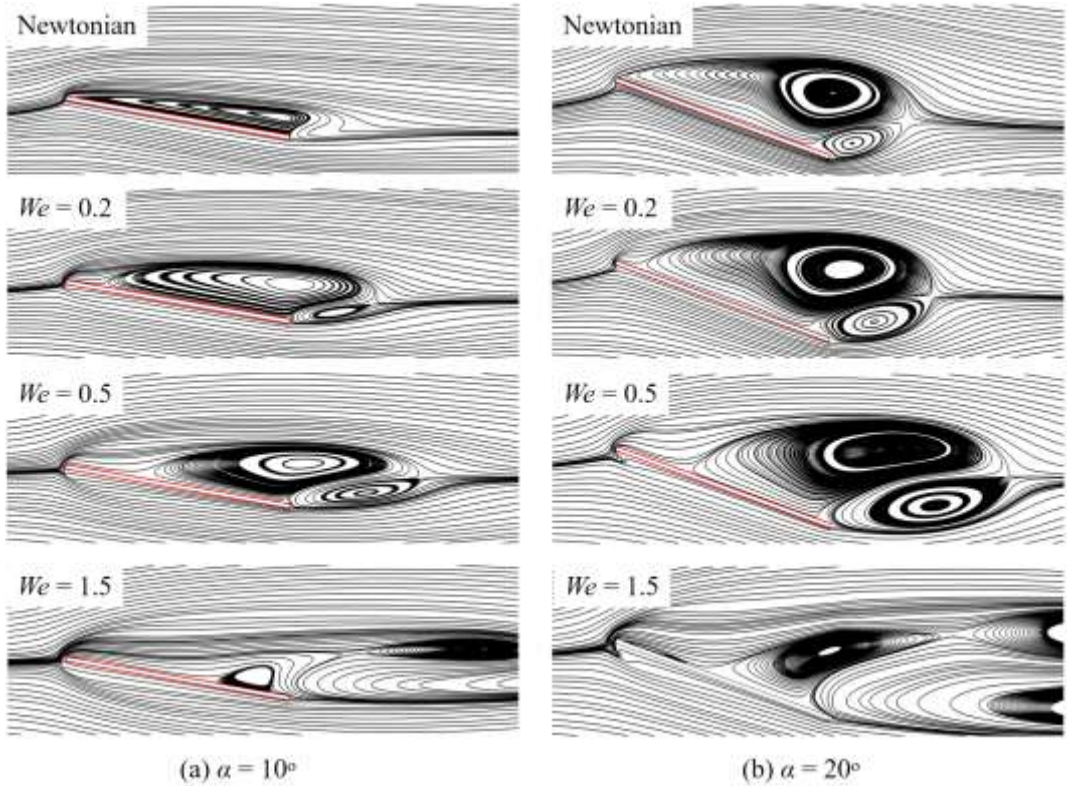

**Fig. 6.** The time-averaged streamlines at different *We* for (a) $\alpha = 10°$ and (b) $\alpha = 20°$.

The drag coefficient ($C_d$) and the lift coefficient ($C_l$) are expressed as,

$$C_d = \frac{2F_x}{\rho U_\infty^2 a} \quad \text{and} \quad C_l = \frac{2F_y}{\rho U_\infty^2 a}, \tag{5}$$

where $F_x$ and $F_y$ are the forces acting on the inclined flat plate in the $x$ and $y$ directions, respectively. The method to calculate the forces could refer to our previous publications.[15,32,33] Besides the time-averaged drag and lift coefficients ($\overline{C_d}$ and $\overline{C_l}$), the root mean square of drag and lift coefficients ($C_{drms}$ and $C_{lrms}$) and the Strouhal number ($St = fa/U_\infty$, where $f$ is the vortex shedding frequency and obtained by fast Fourier transform (FFT) of lift coefficient time series) are also considered.

A grid independence study is performed by smoothly halving and doubling the present grid points in both the $x$ and $y$ directions. Mesh1, Mesh2 and Mesh3 denote the coarse mesh, the present mesh, and the dense mesh, respectively. The simulation case considered here is viscoelastic flow over an inclined flat plate for $\alpha = 20°$, $Re = 500$ and $We = 1.5$. The effect of the mesh sizes on the time-averaged lift coefficient ($\overline{C}_l$) is presented in Table 1, which indicates that the present mesh (Mesh2) size is fine enough to obtain the grid independence results.

To validate our simulation for this special geometry, we compare our numerical results with those of Yang et al.[60] for Newtonian fluid at $Re = 500$ and $\alpha = 20°$, as listed in Table 2. The root mean square values of drag and lift coefficients are indeed quite different from the results of Yang et al.[60] with the difference being more than 20%. This may be due to the numerical calculation method, mesh, statistical strategy and other factors. However, the time-averaged statistical results coincide well with Yang et al.[60] For example, the time-averaged lift coefficient ($\overline{C}_l$) is 0.9814 in the present study and 0.9823 in Yang et al.[60] For viscoelastic flow, variation of $\overline{C}_l$ with $We$ is plotted in Fig. 5. $\overline{C}_l$ gradually decreases with an increase in $We$ for both $\alpha = 10°$ and $20°$. The magnitude of lift could reflect the degree of asymmetric distribution of fluid's force on the flat plate upper and lower surfaces. Decreasing $\overline{C}_l$ indicates that the flow asymmetry is weakened near the inclined flat plate wall.

**Table 1.** Grid independence study.

| Mesh | Mesh1 | Mesh2 | Mesh3 |
|------|-------|-------|-------|
| $\overline{C}_l$ | 0.5377 | 0.5075 | 0.5067 |

**Table 2**. Comparison of force coefficients at $Re = 500$ and $\alpha = 20°$ for Newtonian fluid.

|  | $\overline{C}_d$ | $\overline{C}_l$ | $St$ | $C_{lrms}$ | $C_{drms}$ |
|------|------|------|------|------|------|
| Yang et al.[60] | 0.4472 | 0.9823 | 0.4959 | 0.1003 | 0.0176 |
| Present Results | 0.4716 | 0.9814 | 0.4923 | 0.1262 | 0.0144 |

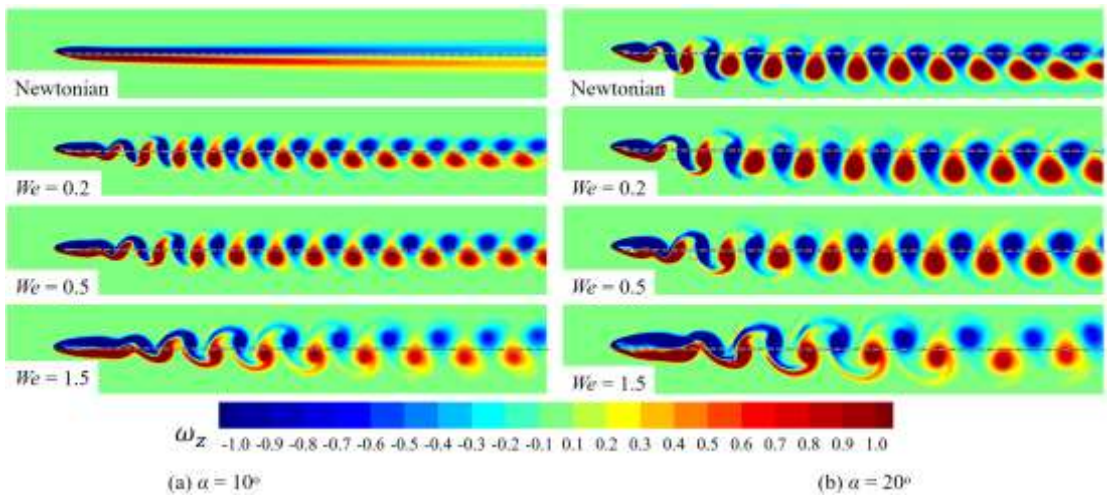

**Fig. 7.** The instantaneous vorticity contours at different $We$ for (a) $\alpha = 10°$ and (b) $\alpha = 20°$. $\omega_z$ is normalized by $U_\infty/a$.

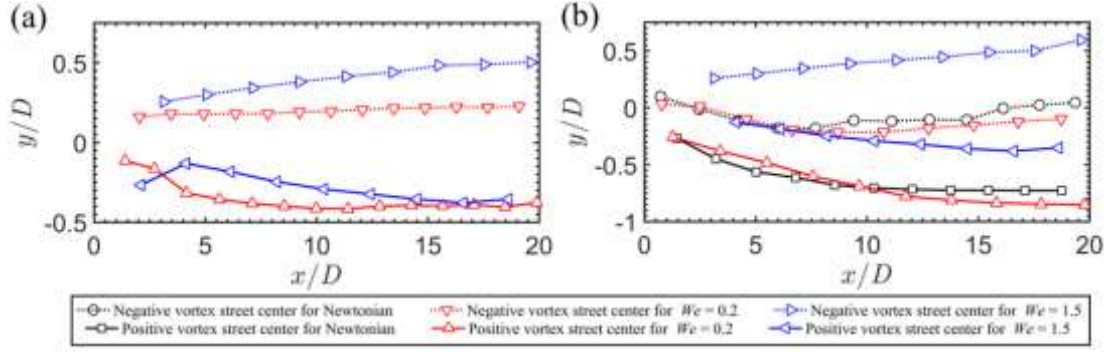

**Fig. 8.** The corresponding instantaneous positions of vortex cores for Newtonian and viscoelastic flows at (a) α = 10° and (b) α = 20° shown in Fig. 7.

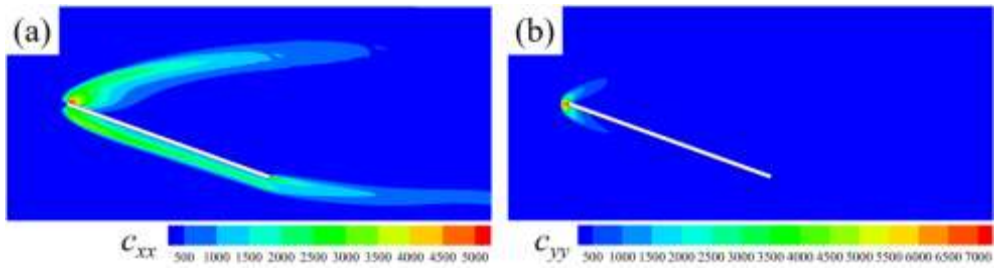

**Fig. 9.** The instantaneous distributions of (a) $c_{xx}$ and (b) $c_{yy}$ for the viscoelastic flow over an inclined flat plate at α = 20° and We = 1.5.

The time-averaged streamlines at different *We* for α = 10° and 20° are shown in Fig. 6. The recirculation region behind the flat plate gradually elongates with increasing *We*. The corresponding instantaneous vorticity contours are shown in Fig. 7. In Newtonian fluid, at high α (e.g. α = 20°), the vortex shedding trajectory deviates from the horizontal center line of the computational domain. However, as *We* increases, the vortex shedding trajectory gradually approaches to the horizontal center line. The corresponding instantaneous positions of vortex cores for Newtonian and viscoelastic fluids at α = 10° and 20° are extracted and plotted in Fig. 8. Clearly, with an increase in *We*, the vortices on both sides of the flat plate tend to be symmetric against the horizontal center line. This trend is similar to that for the viscoelastic flow over a NACA0012 with a certain angle of attack reported in Xiong *et al.*[32]

For viscoelastic flow over an inclined flat plate with various angles of incidence, a region with high elastic stress appears around the leading edge and then extends to the wake region (refer to the regions of high conformation tensor components $c_{xx}$ and $c_{yy}$ in Fig. 9(a) and (b)). α does not fundamentally affect the distribution of elastic stress (data not shown). The distribution of the corresponding time-averaged flow velocity magnitude ($\sqrt{\bar{u}^2 + \bar{v}^2}/U_\infty$) is shown in Fig. 10(b), which is compared with that of Newtonian fluid shown in Fig. 10(a). The contour of $\sqrt{\bar{u}^2 + \bar{v}^2}/U_\infty = 0.1$ is marked with the pink line in Fig. 10(b). The flow velocity within the region enclosed by the pink line is lower than 0.1, which could be regarded as a 'dead zone'. The 'dead zone' tends to be symmetric with the horizontal center line (the dark dash line) with increasing *We*. The 'dead zone' is surrounded by the high elastic stress region shown in Fig. 9(a) and (b). The high elastic region resembles a symmetric shield, and thus the vortex shedding trajectory also tends to be symmetric in its wake flow.

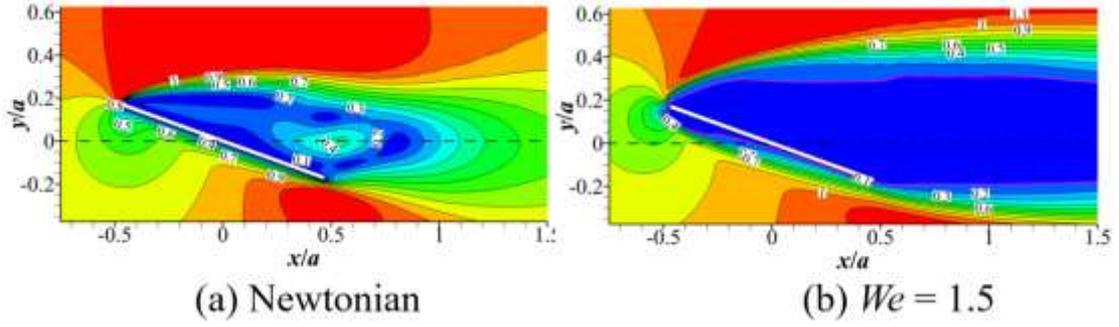

**Fig. 10.** The distribution of time-averaged flow velocity magnitude ($\sqrt{\bar{u}^2 + \bar{v}^2}/U_\infty$) for (a) Newtonian fluid and (b) viscoelastic fluid with $We = 1.5$ and $L = 100$. The contour of $\sqrt{\bar{u}^2 + \bar{v}^2}/U_\infty = 0.1$ is marked with the pink line in panel (b).

## B. Viscoelastic flow over a rotating circular cylinder

Second, we consider viscoelastic flow over a rotating circular cylinder in a free domain, as shown in Fig. 11(a). The whole computational domain has a rectangular shape, with the length $L_u + L_d$ and the width $H$. The center of the cylinder is set at $(x, y) = (0, 0)$. The diameter of the cylinder is $D$. The distance between the inlet and the cylinder center is set as $L_u = 25D$. The distance between the outlet and the cylinder center is set as $L_d = 75D$. $H$ is set as $50D$. The free stream boundary condition with the uniform velocity $u = (U_\infty, 0)$ is imposed on the inlet. The no-slip boundary condition is adopted on the cylinder surface. The angular velocity of rotating cylinder is $\omega$ and the dimensionless rotation velocity rate is defined as $\alpha_r = \omega D/(2U_\infty)$. In this study, $\alpha_r$ ranges from 0 to 6. A slip condition is imposed on the upper and lower boundaries of the computational domain. The pressure at the outlet is set as $p = 0$. For the conformation tensor, the no-flux condition is approximated on all boundaries except for the inlet boundary with $\mathbf{c} = \mathbf{I}$. The Reynolds number is defined as $Re = U_\infty D\rho/\mu$ and fixed at 100 and the Weissenberg number is defined as $We = \lambda U_\infty/D$, which ranges from 0 to 10.

As shown in Fig. 11(b), the surrounding region of the circular cylinder is discretized by the O-type mesh. The rest of the computational domain is discretized by several blocks of quadrilateral meshes, with the dense mesh near the cylinder and the coarse mesh near the domain boundaries. The O-type mesh consists of 280 grid points uniformly distributed along the cylinder perimeter and 71 grid points stretched with an exponential progression (exponential ratio of 1.02) along the radial direction to ensure a fine mesh near the cylinder surface. In this study, the height of the first cell adjacent to the cylinder surface in the radial direction is set to $0.0025D$. In the x-direction, 501 grid points (for $L_v$) are unevenly arranged in the downstream region and 37 grid points (for $L_u$) are set in the upstream region. In the y-direction, the upper and lower regions (separated by the center region) both have 51 unevenly distributed grid points. The total number of meshes for the computational domain is approximately 85, 000. The time step is set to be small enough in our simulation to ensure the stability of numerical calculations. For most simulation cases, the time step is set to $0.005D/U_\infty$. For the certain cases that are more difficult to converge (mainly those at high $We$, e.g., $We > 2$), the calculation time step is reduced to $0.0025D/U_\infty$ or even $0.00125D/U_\infty$.

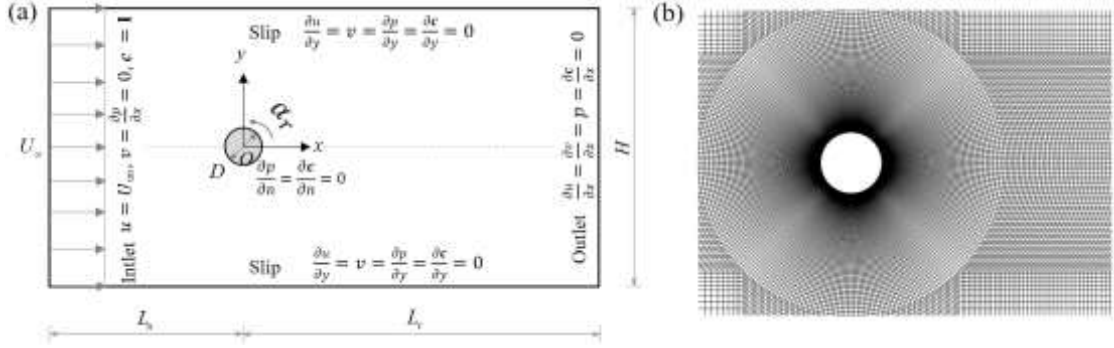

**Fig. 11.** (a) The schematic of flow over a rotating circular cylinder. (b) The mesh topology near the cylinder.

The lift coefficient ($C_l$) is calculated as,

$$C_l = \frac{2F_y}{\rho U_\infty^2 D},$$ (6)

where $F_y$ is the force acted on the cylinder in $y$ direction.

The absolute time-averaged lift coefficients ($|\overline{C_l}|$) of the cylinder is calculated and plotted in Fig. 12. Fig. 12(a) shows that $|\overline{C_l}|$ increases with $\alpha_r$ in Newtonian fluid, which agrees well with those reported by Bourguet & Jacono[61] and Stojković *et al.*[9] However, all the numerical results show obvious difference with the potential flow solution[62] (the dotted line, $|\overline{C_l}| = 2\pi\alpha_r$), due to flow separation and vortex shedding induced by the viscous effect. In viscoelastic fluid, $|\overline{C_l}|$ gradually decreases with an increase in *We* for all $\alpha_r$ as shown in Fig. 12(b). The magnitude of lift could reflect the degree of asymmetric distribution of the hydrodynamic force on the upper and lower parts of cylinder surface. At high *We*, such as *We* = 10, $|\overline{C_l}|$ increase a little as *We* increases. For viscoelastic flow, the lift coefficient contains three components, i.e., pressure contribution $\overline{C_l^{pressure}}$, viscous contribution $\overline{C_l^{viscous}}$, and polymer contribution $\overline{C_l^{polymer}}$. Fig. 13 shows the variation of the lift coefficient component with *We* at $\alpha_r = 1$, which indicates that the pressure contribution is dominant in the lift coefficient. Thus, Fig. 14 displays the time-averaged pressure distributions for *We* = 0.2, *We* = 2 and *We* = 10 at $\alpha_r = 1$. The pressure coefficients for the upper and lower halves of the cylinder surface are also calculated and marked in Fig. 14. It can be seen that the local time-averaged pressure along the whole cylinder becomes higher with increasing *We*. For low *We* range of *We* ≤ 2, the pressure increase along the lower half surface is more obvious than that along the upper half surface. Thus, $|\overline{C_l}|$ increases with *We*. However, when further increasing *We*, the pressure along the upper half surface increases more than that along the lower half surface, which eventually results in a decrease in $|\overline{C_l}|$ with a further increase in *We*.

The instantaneous vorticity contours (left column) and the time-averaged streamlines (right column) at (*Re*, $\alpha_r$, *L*) = (100, 1.8, 100) for different *We* are shown in Fig. 15. The corresponding instantaneous positions of vortex cores are extracted and drawn in Fig. 16. At this $\alpha_r$, in Newtonian fluid, the vortex shedding exhibits the feature of mode I (the same to $\alpha_r = 0$). However, the vortex shedding trajectory deviates from the horizontal center line of the cylinder (the dotted line). In viscoelastic flow, the vortex shedding trajectory gradually approaches the horizontal

Wake asymmetry weakening in viscoelastic fluids center line with increasing *We*. At the same time, the time-averaged recirculation wake elongates and tends to be symmetric with respect to the center line. A rotating boundary layer with circular streamlines forms surrounding the cylinder, which grows with increasing *We*. This boundary layer region separates the time-averaged recirculation wake region with the cylinder.

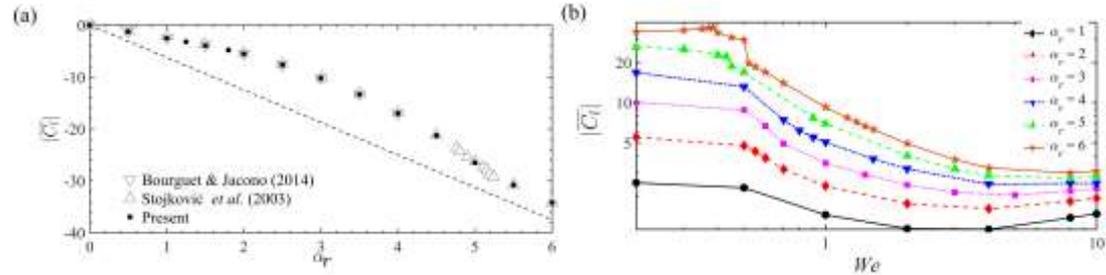

**Fig. 12.** The absolute time-averaged lift coefficients of the rotating cylinder for (a) Newtonian fluid and (b) viscoelastic fluid. The dotted line denotes the potential flow solution.

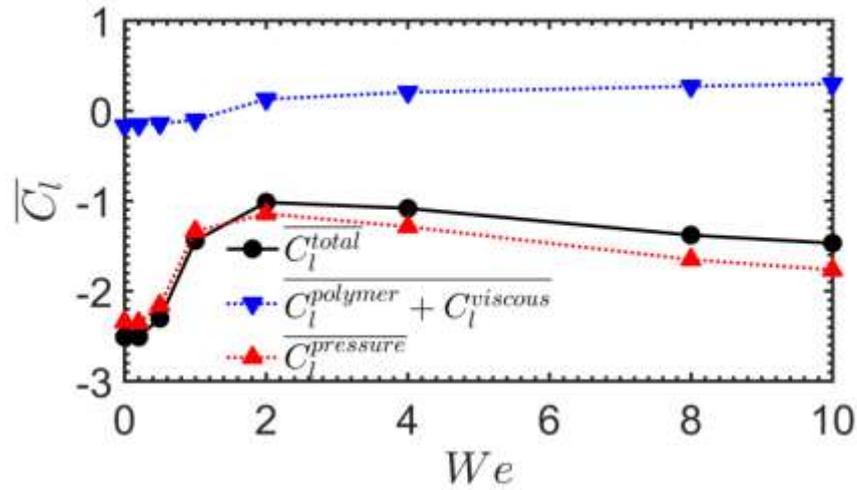

**Fig. 13.** Variation of the time-averaged lift coefficient components with *We* for $\alpha_r = 1$.

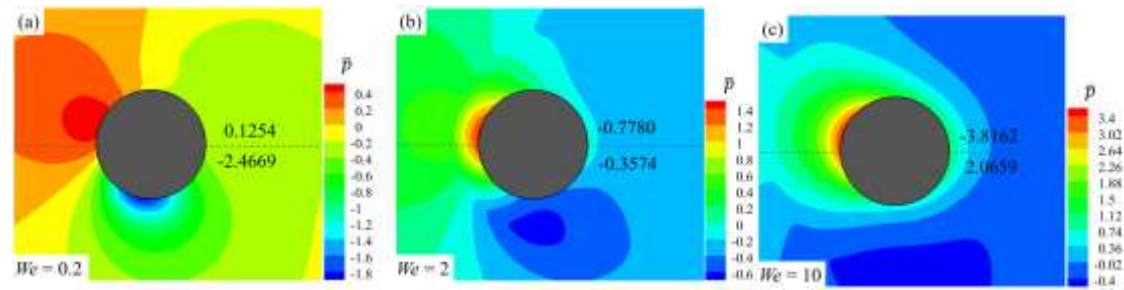

**Fig. 14.** The time-averaged pressure distribution at $\alpha_r = 1$ for (a) *We* = 0.2, (b) *We* = 2, and (c) *We* = 10. The numbers upper and below the dotted line denote the pressure coefficients for the upper and lower halves of the cylinder surface, respectively.

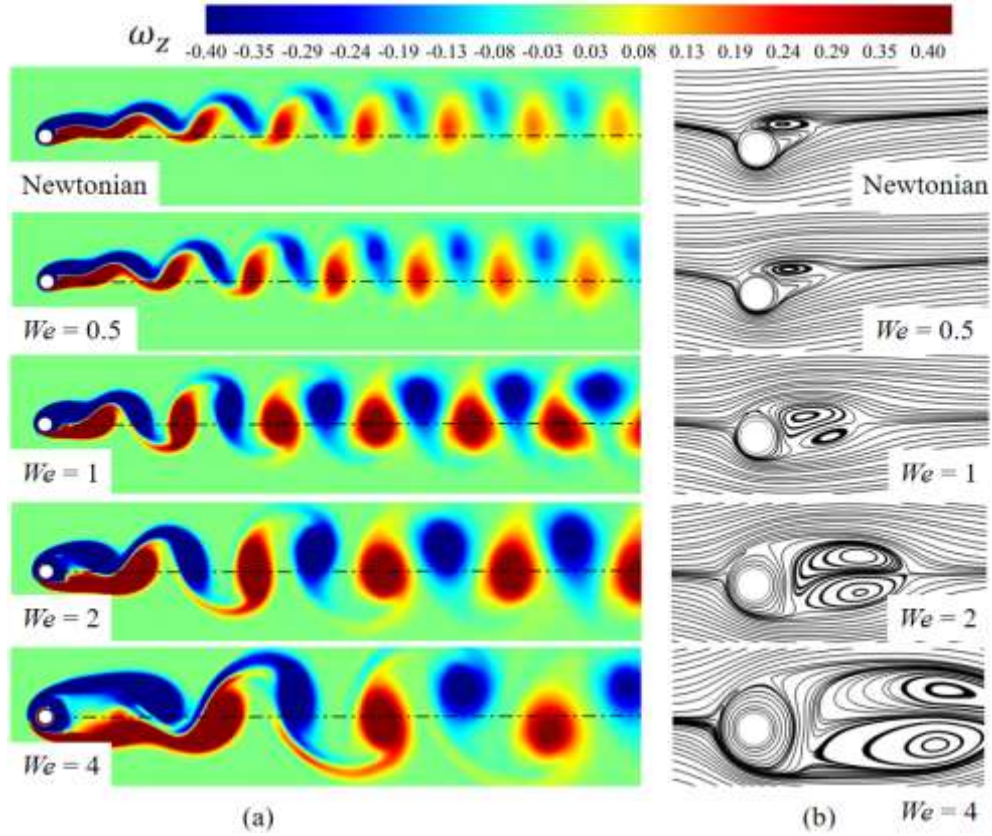

(a)                                                      (b)

**Fig. 15.** (a) The instantaneous vorticity contours and (b) the time-averaged streamlines for $\alpha_r = 1.8$. $\omega_z$ is normalized by $U_\infty/D$.

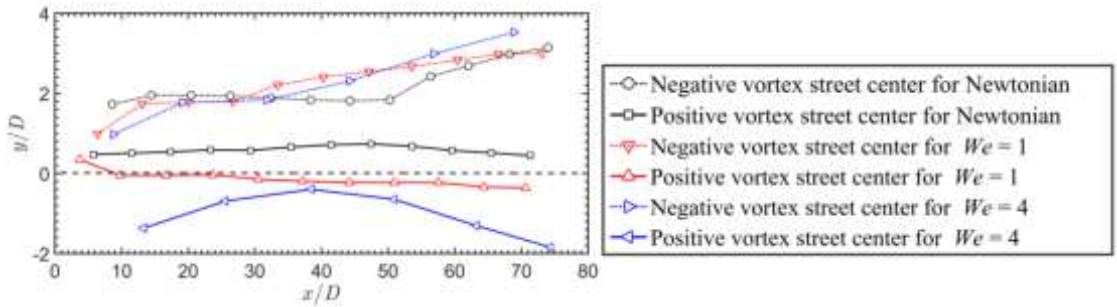

**Fig. 16.** The corresponding instantaneous positions of vortex cores for Newtonian and viscoelastic flows with $\alpha_r = 1.8$ shown in Fig. 15(a).

The instantaneous vorticity contours (left column) and the time-averaged streamlines (right column) at $(Re, \alpha_r, L) = (100, 6.0, 100)$ for different $We$ are shown in Fig. 17. At this $\alpha_r$, in Newtonian fluid, the flow is steady. Due to the strong rotation effect, the time-averaged streamlines form an egg shape, surrounding the cylinder. In viscoelastic flow at $We = 1$, the flow is still steady. However, the vorticity generated near the cylinder surface extends downstream. For the time-average flow streamlines, the 'egg' shape formed in Newtonian fluid now becomes an 'annulus' surrounding the cylinder. At the same time, another recirculation region, whose location is determined by the combination effect of the free stream and rotating cylinder, appears downstream and attaches to the outer surface of the annulus. When *We* increases to 2, the flow becomes unsteady. The vortex sheds downstream, which deviates from the horizontal center line of the cylinder (the dash-dotted lines in Fig. 17). The time-averaged flow streamlines at *We* = 2 show that both the 'annulus' and the downstream recirculation region become larger. In viscoelastic flow at *We* = 4, the vortex shedding trajectory approaches the horizontal center line. The time-averaged flow streamlines indicate continuous growth of the 'annulus' and the downstream recirculation region. The downstream recirculation region tends to be symmetric with respect to the horizontal center line.

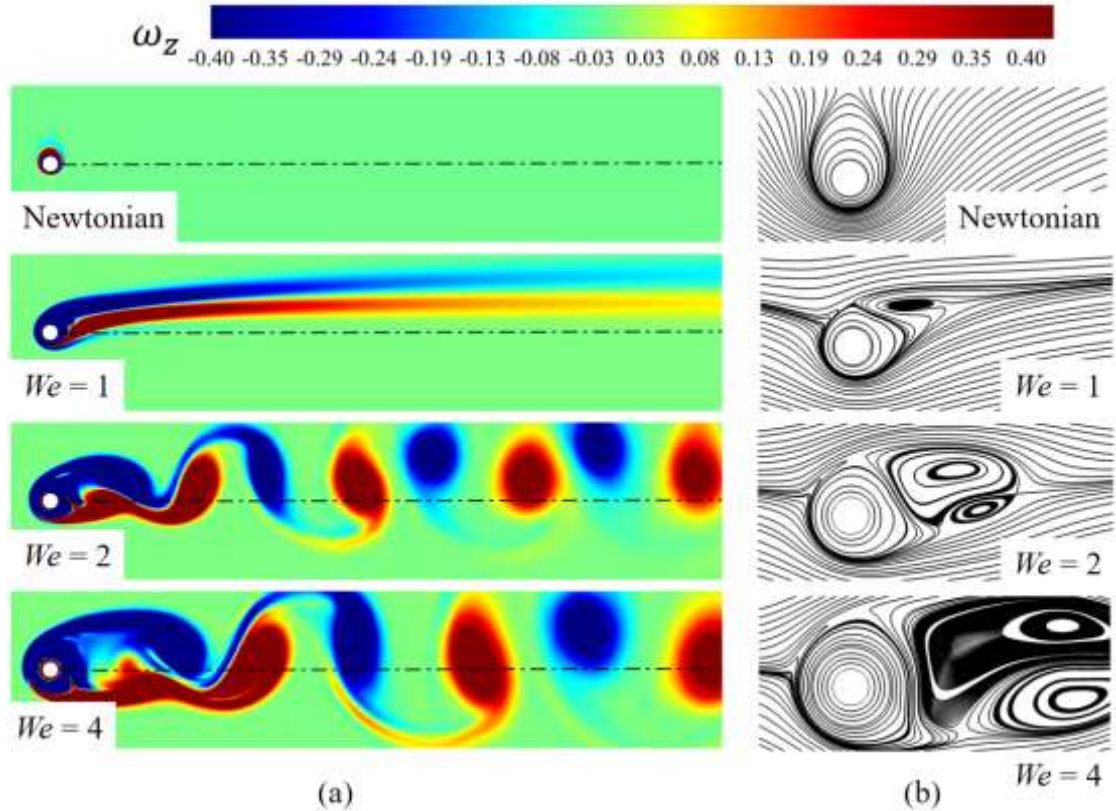

**Fig. 17.** (a) The instantaneous vorticity contours and (b) the time-averaged streamlines for $\alpha_r$ = 6.0. $\omega_z$ is normalized by $U_\infty/D$.

When *We* is high (no less than 4), the flow behaves similar, as shown before for $\alpha_r$ = 1.8 and $\alpha_r$ = 6.0. Then, we consider a higher *We* of 10. The instantaneous vorticity contours (left column) and the time-averaged streamlines (right column) at *We* = 10 for different $\alpha_r$ are shown in Fig. 18. The flow all consists of two parts, i.e., a thick rotating boundary layer flow around the rotating circular cylinder and vortex shedding in the wake. A simple model, as shown in Fig. 19, could be used to describe the time-average flow behavior of the viscoelastic fluid past a rotating circular cylinder at a high *We* (such as *We* = 10). $D_\alpha$ denotes the thickness of the rotating boundary layer ($D_\alpha > D$), which increases with $\alpha_r$. We notice that the profile of the circumferential velocity $u_\theta$ along the radial direction *r* in the rotating boundary layer is almost completely consistent, as shown in Fig. 19. $u_\theta$ gradually decays with *r* and tends to zero at the outer edge of the rotating boundary layer $r = D_\alpha/2$. Thus, this outer edge $r = D_\alpha/2$ acts as a dummy stationary solid wall and divides the flow into two regimes, as shown in Fig. 20. The flow induced by the rotating cylinder is shielded by this dummy wall and does not communicate with the free stream outside. On the other hand, the free stream outside resembles passing a larger stationary cylinder with $r = D_o/2$ and the wake becomes symmetric.

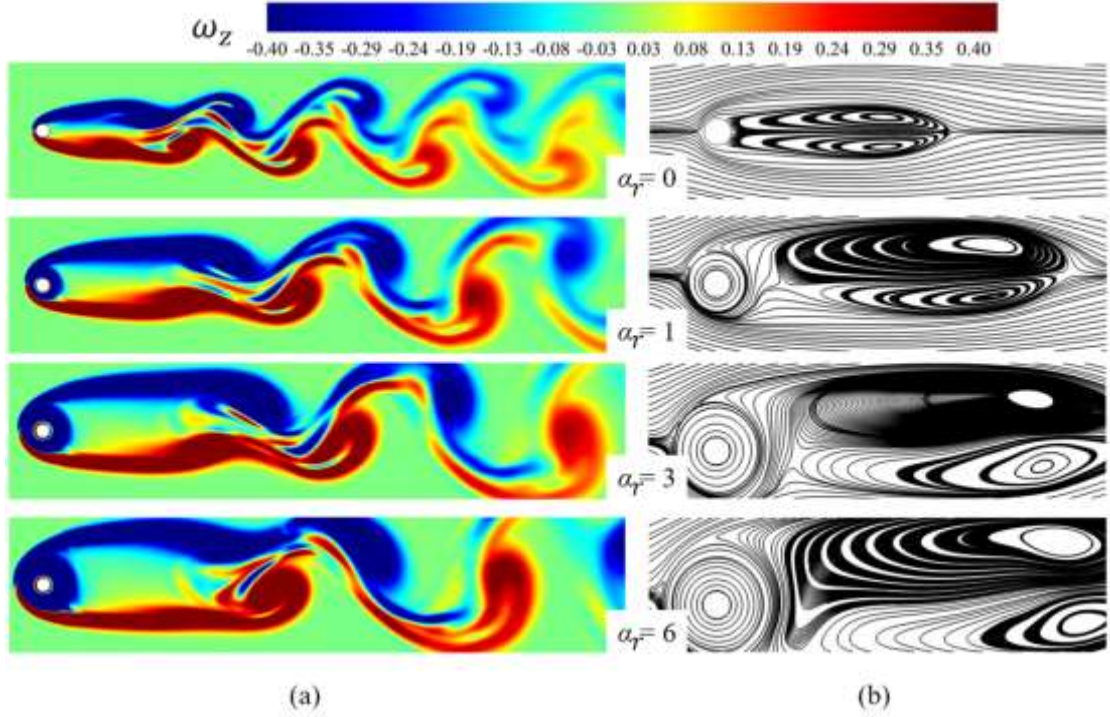

**Fig. 18.** The instantaneous vorticity contours (left column) and the time-averaged streamlines (right column) at $(Re, We, L) = (100, 10, 100)$. $\omega_z$ is normalized by $U_\infty/D$.

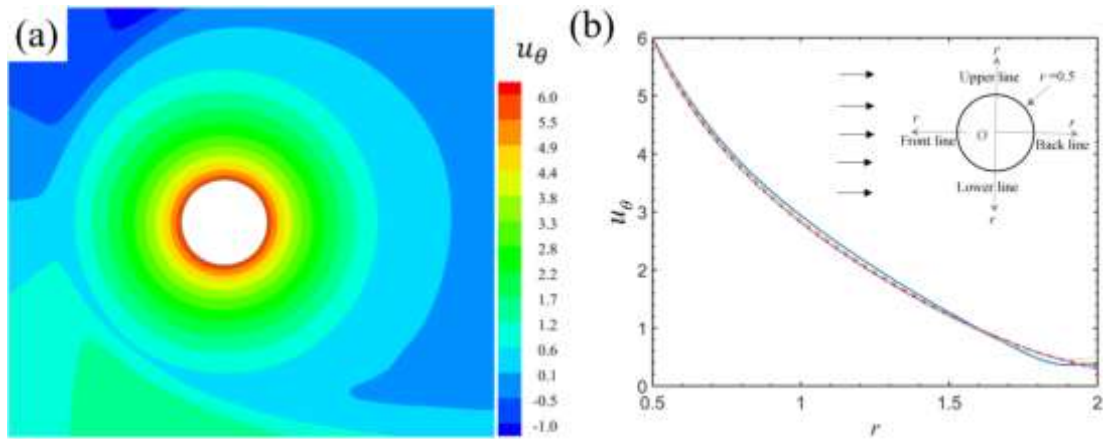

**Fig. 19.** (a) The instantaneous distributions of the circumferential velocity $u_\theta$. (b) Circumferential velocity ($u_\theta$) profiles along the front, back, upper and lower lines near the cylinder for $(Re, We, L, \alpha_r) = (100, 10, 100, 6)$. $r = 0.5$ denotes the cylinder surface.

Wake asymmetry weakening in viscoelastic fluids

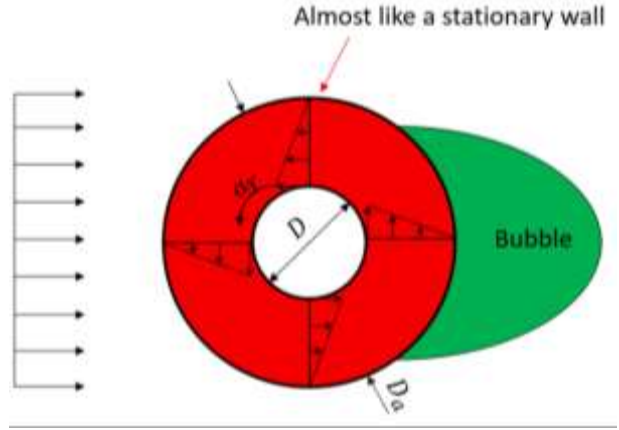

**Fig. 20.** A simple model to illustrate the feature of the viscoelastic flow over a circular cylinder rotating at a high *We*.

In order to understand the above-mentioned velocity distribution within the rotating boundary layer, the elastic stress distribution is considered. The distributions of the conformation tensor components ($c_{xx}$ and $c_{yy}$) are shown in Fig. 21. Conformation tensor is positively related to elastic stress. Two elastic stress boundary layers can be identified by combining the regions of high conformation tensor components $c_{xx}$ and $c_{yy}$ in Fig. 21(a) and (b). One is a thick inner elastic stress boundary layer surrounding the rotating cylinder, which is caused by the velocity gradient induced by the rotation of the cylinder. Another is an outer elastic stress boundary layer surrounds the dummy stationary solid cylinder ($D_\alpha$). In viscoelastic fluid, elastic stress usually tends to appear in the place where the velocity gradient is large. Conversely, the existence of elastic stress gradient plays a dissipative role, which in turn weakens the gradient of velocity. These two effects compensate each other, which result in a uniform tangential velocity at a specific radial location with respect to the center of the cylinder. This uniform tangential velocity decays along the radial direction and becomes nearly zero on the outer boundary of this elastic stress boundary layer. Thus, the external flow resembles that past a cylinder with larger diameter $D_\alpha$ and exhibits a symmetric vortex shedding trajectory.

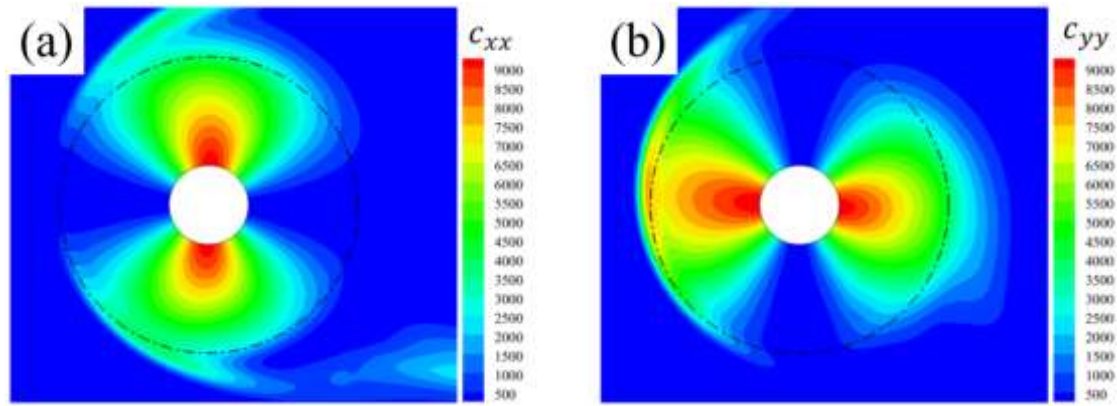

**Fig. 21.** The instantaneous distributions of (a) $c_{xx}$ and (b) $c_{yy}$ for the viscoelastic flow over a rotating circular cylinder at $\alpha_r = 6$ and $We = 10$. The dash-dotted lines denote the position of the dummy stationary solid wall.

### C. Viscoelastic flow over a circular cylinder with asymmetric slip boundary distribution

Third, we consider viscoelastic flow over a circular cylinder with asymmetric slip boundary distribution in a free domain, as shown in Fig. 22(a). All the boundary conditions are the same as those in the previous example of the rotating cylinder except the velocity boundary condition for the cylinder surface. A partial slip boundary condition is imposed on the upper half of the cylinder surface while the non-slip boundary condition is imposed on the lower half. The mesh is generated by the commercial software ICEM, as shown in Fig. 22(b). In particular, the O-type mesh in the vicinity of the cylinder consists of 300 grid points along the cylinder perimeter (with dense distribution near the discontinuous point between the partial slip and non-slip boundary condition) and 71 grid points stretched exponentially along the radial direction. To minimize the discontinuity effect of the velocity boundary condition, the grid line is purposely located at the junction of the partial slip and non-slip boundaries. The size of the first cell adjacent to the cylinder surface in the radial direction is set to $0.000125D$. In the $x$-direction, 501 grid points (for $L_v$) are unevenly arranged in the downstream region and 71 grid points (for $L_u$) are set in the upstream region. The total number of meshes for the computational domain is approximately 300,200. A preliminary test indicates that the present mesh resolution is fine enough to obtain the grid independence results. For all simulation cases, the time step is set to $0.0125D/U_\infty$. The time step is small enough to ensure the stability and accuracy of numerical calculations. In order to minimize the effect of artificial dissipation, the artificial dissipation is set to nonzero only for the cases with $We > 0.35$.

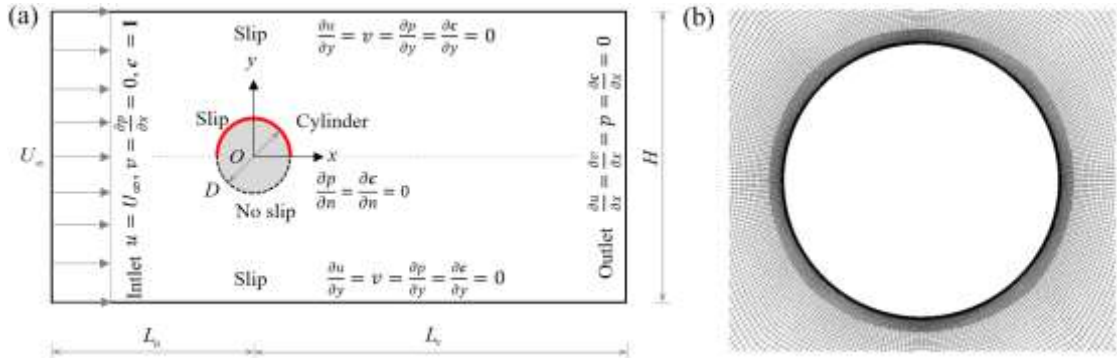

**Fig. 22.** (a) The schematic of the unconfined flow around a circular cylinder with asymmetric slip boundary distribution. (b) The mesh topology near the cylinder.

In this example, the Reynolds number is defined as $Re = U_\infty D\rho/\mu$ and fixed at 500 and the Weissenberg number is defined as $We = \lambda U_\infty/D$, which ranges from 0 to 1.0. The linear slip length ($L_s$) method is adopted to describe the degree of the cylinder's upper wall slippage. The slip length is defined as,[63]

$$\mathbf{n} \times \mathbf{u} = L_s \mathbf{n} \times [(\nabla \mathbf{u} + (\nabla \mathbf{u})^{\mathrm{T}}) \cdot \mathbf{n}], \tag{7}$$

where $\mathbf{n}$ is the unit normal to the cylinder surface. A dimensionless number, Knudsen number ($Kn$) is used to evaluate the relative relationship between the slip length and the cylinder diameter, which reads,

$$Kn = L_s/D. \tag{8}$$

According to Legendre *et al.*,[64] Eq. (7) is calculated by using the velocity on the two-layer grids immediately close to the cylinder surface. In this simulation, $Kn = 0.5$ is adopted for the upper half of the cylinder surface.

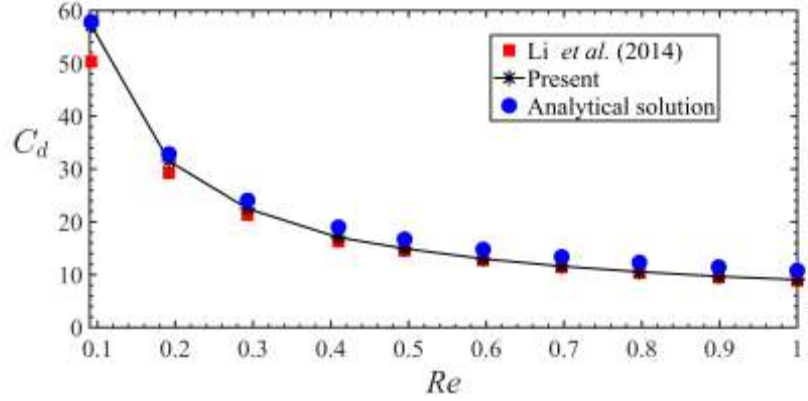

**Fig. 23.** Variation of drag coefficients ($C_d$) with $Re$ in Newtonian fluid. The partial slip boundary condition with $Kn = 1$ is imposed on the whole cylinder surface.

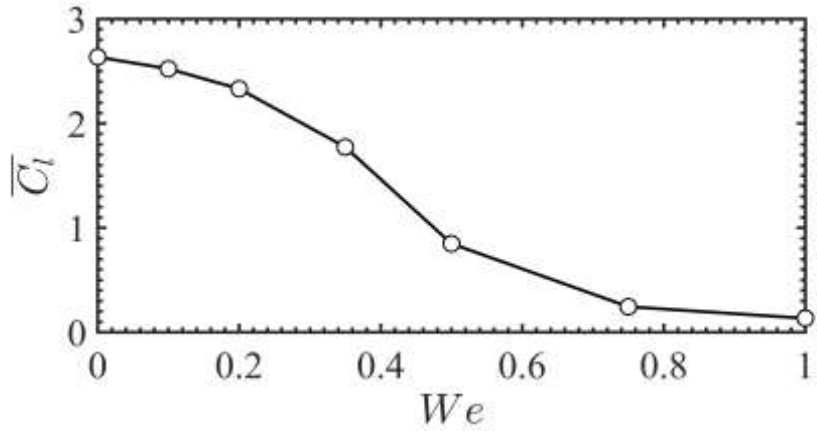

**Fig. 24.** The time-averaged lift coefficients for viscoelastic flow over a circular cylinder with the asymmetric slip boundary condition at $Re = 500$ and $Kn = 0.5$.

Li *et al.*[10] deducted an approximate analytical solution for laminar Newtonian flow over a circular cylinder with the partial slip boundary on the whole cylinder surface. The corresponding $C_d$ can be calculated by,

$$C_d = \left[ \frac{1}{1 - \dfrac{1}{2(1 + 2Kn)} - \gamma - \ln(Re/8)} \right] \cdot \frac{8\pi}{Re},$$ (9)

where $\gamma$ is the Euler constant. Eq. (9) is more accurate when $Re$ is much less than 1. We also simulate this flow at a fixed $Kn$ of 1 and over the range of $Re$ from 0.1 to 1. At this range of $Re$, a very low block ratio ($BR$, the ratio of the cylinder diameter $D$ to the width of the computational domain $H$) has to be adopted to avoid the effect of the lateral boundaries. In our simulation, $BR$ is set as 1‰. Our simulated $C_d$ are compared well with the analytical solution and the simulation results of Li *et al.*[10] ($BR = 1\%$), as shown in Fig. 23. When $Re$ is less than 0.2, our results almost coincide with the analytical solution, but obviously different from those of Li *et al.*[10] The difference may result from the different $BR$ because the effect of $BR$ on $C_d$ is more obvious at low $Re$. When $Re$ is higher than 0.3, our results compare well with those of Li *et al.*[10] This validation confirms that the present method is capable to capture the flow feature induced by the partial slip boundary condition.

The time-averaged lift coefficient ($\overline{C}_l$) of the cylinder with the asymmetric slip boundary condition at various $We$ is plotted in Fig. 24. In Newtonian fluid, $\overline{C}_l$ is 2.665. In viscoelastic fluid, $\overline{C}_l$ decreases with an increase in $We$. The magnitude of lift could reflect the degree of the asymmetric distribution of hydrodynamic force along the cylinder surface. The decreasing $\overline{C}_l$ indicates that the flow asymmetry is weakened near the cylinder surface when $We$ is sufficiently large. In the following, the wake flow behavior in viscoelastic fluid is discussed.

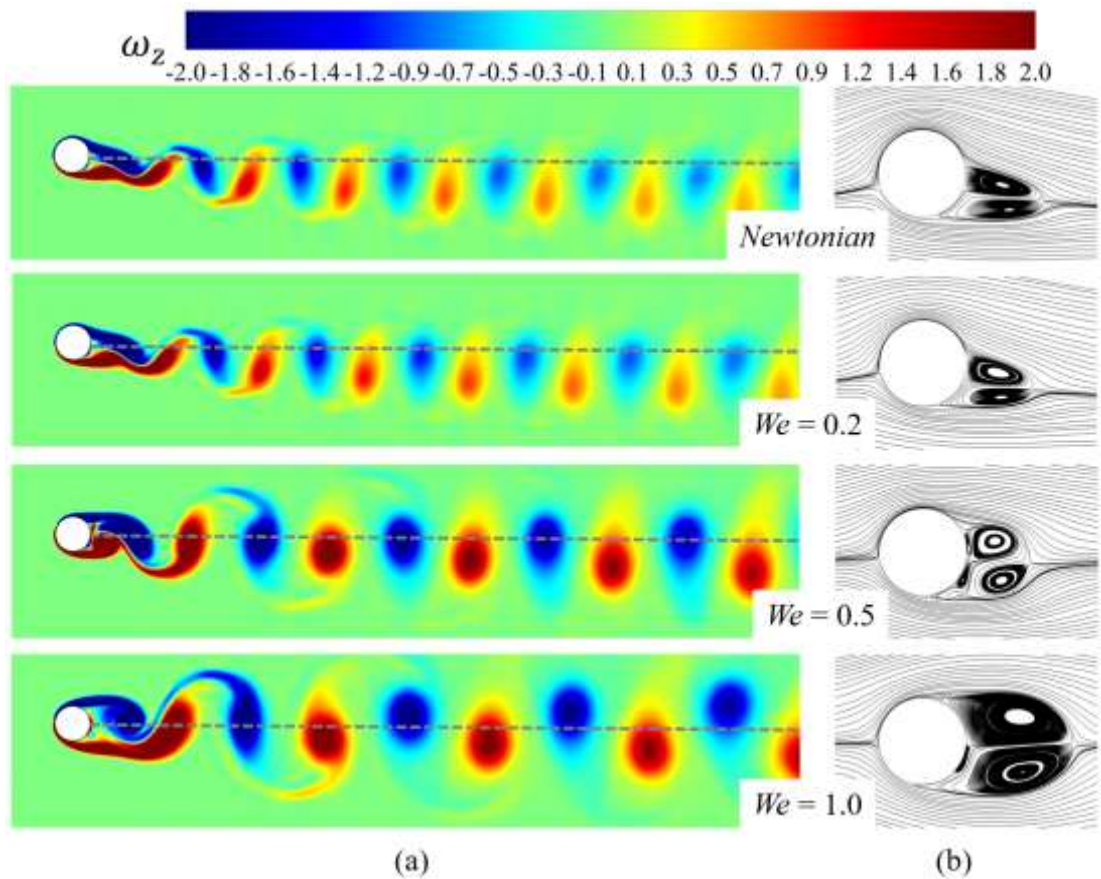

**Fig. 25.** (a) The instantaneous vorticity contours and (b) the time-averaged streamlines for different $We$. $\omega_z$ is normalized by $U_\infty/D$.

Wake asymmetry weakening in viscoelastic fluids

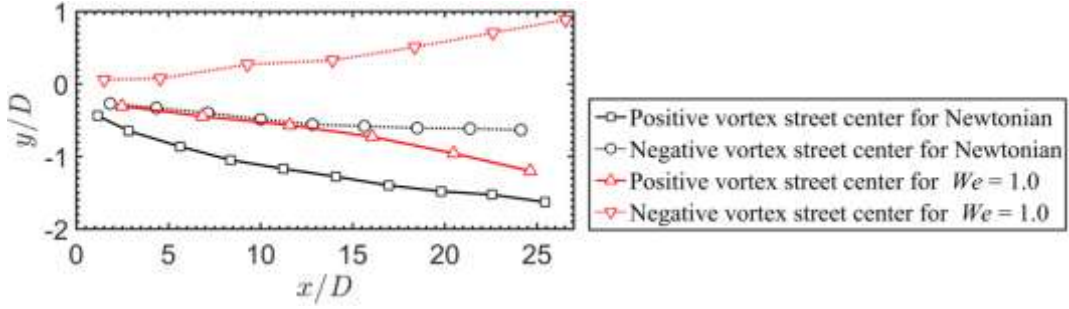

**Fig. 26.** The corresponding instantaneous positions of vortex cores for Newtonian and viscoelastic flows shown in Fig. 25(a).

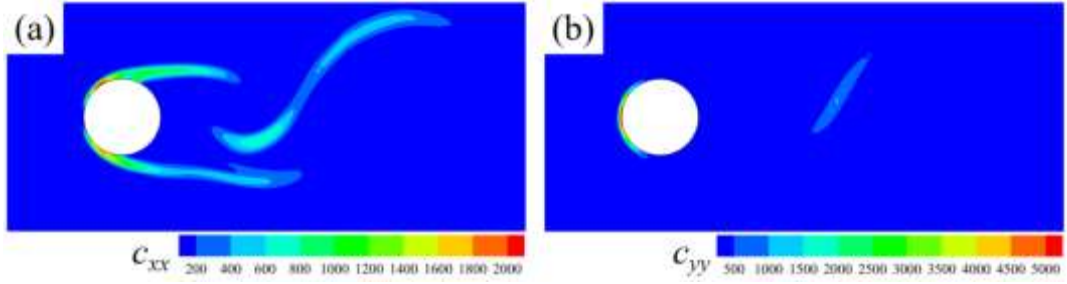

**Fig. 27.** The instantaneous distributions of (a) $c_{xx}$ and (b) $c_{yy}$.

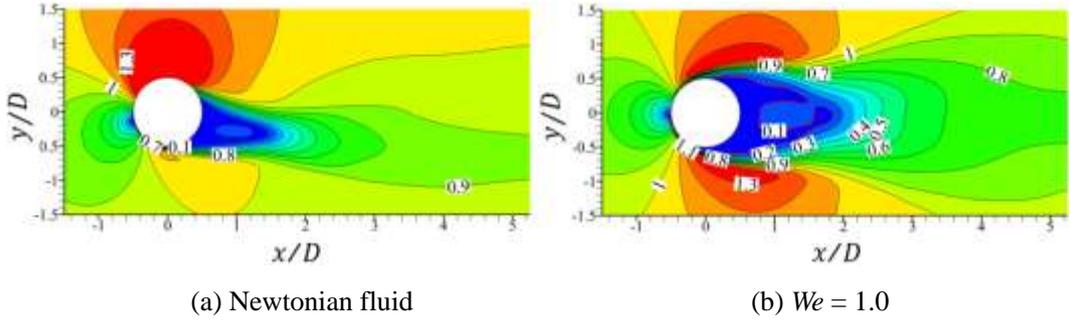

(a) Newtonian fluid  (b) $We = 1.0$

**Fig. 28.** The distribution of time-averaged flow velocity magnitude ($\sqrt{\bar{u}^2 + \bar{v}^2}/U_\infty$) for (a) Newtonian fluid and (b) viscoelastic fluid with $We = 1.0$ and $L = 100$. The contour of $\sqrt{\bar{u}^2 + \bar{v}^2}/U_\infty = 0.1$ is marked with the pink line in panel (b).

The instantaneous vorticity contours (left column) and the time-averaged streamlines (right column) at ($Re$, $Kn$, $L$) = (100, 0.5, 100) for different $We$ are shown in Fig. 25. The corresponding instantaneous positions of vortex cores are extracted and shown in Fig. 26. In Newtonian fluid, the vortex shedding trajectory deviates from the horizontal center line of the cylinder (the dash line in Fig. 25). In viscoelastic fluid, the vortex shedding trajectory gradually approaches to the center line as $We$ increases. At the same time, the time-averaged recirculation wake is elongated and tends to be symmetric against the center line.

For viscoelastic flow over a circular cylinder with the asymmetric slip boundary condition, a region with high elastic stress appears around the leading edge and then extends to the wake region (refer to the regions of high conformation tensor components $c_{xx}$ and $c_{yy}$ in Fig. 27(a) and (b)). The existence of high elastic stress affects the velocity distribution near the cylinder surface.

The distribution of the corresponding time-averaged flow velocity magnitude ($\sqrt{\overline{u}^2 + \overline{v}^2}/U_\infty$) is shown in Fig. 28(b), which is compared with that in Newtonian fluid (Fig. 28(a)). In Newtonian fluid, high velocity distributes near upper half cylinder due to the partial slip wall effect, which, however, tends to zero in viscoelastic fluid. The contour of $\sqrt{\overline{u}^2 + \overline{v}^2}/U_\infty = 0.1$ is marked with the pink line in Fig. 28(b). The flow velocity within the region enclosed by the pink line is lower than 0.1. The region with low flow velocity could be regarded as a whole blunt body, which shows symmetric feature. Thus, a symmetric vortex shedding trajectory appears when $We$ is sufficiently large.

### D. Viscoelastic flow over eight equal side-by-side circular cylinders placed inclined closely

Fourth, we consider viscoelastic flow over an inclined row of eight equally closely spaced circular cylinders, as shown in Fig. 29. The cylinders are named from cy-1 to cy-8 from left to right. The inclined angle, i.e., the angle between the cylinder row and the incoming flow direction, is set as $20°$. The diameter of each cylinder is $D$ and the nearest distance between each nearby two cylinders is $L_D = 0.1D$. For Newtonian flow over two side-by-side circular cylinders when $L_D$ is $0.1D$, the vortex shedding behavior resembles that behind a single cylinder.[33] In viscoelastic flow, the wake is also similar to that behind one large cylinder (the cross-sectional area facing the incoming flow is about $2D + L_D D$).[33] However, if $L_D$ increases, such as $L_D = 3D$, the two cylinders may not be considered as a single blunt body. Note that the present study only concerns the wake asymmetry weakening of one blunt body or several closely placed bodies which could be regarded as a single body. Thus, we only consider $L_D = 0.1D$ in this numerical example.

The whole computational domain has a rectangular shape, with the length $L_u + L_v$, and the width $L_w$. The center of the cylinder row is set at the origin of the coordinate system $(x, y) = (0, 0)$. The distance between the inlet and the origin is set as $L_u = 25D$. The distance between the outlet and the origin is set as $L_v = 75D$. $H$ is set as $50D$. The free stream boundary condition with the uniform velocity $\mathbf{u} = (U_\infty, 0)$ is imposed on the inlet. A slip condition is imposed on the two lateral boundaries of the computational domain. The pressure at the outlet is set as $p = 0$. The no-slip boundary condition is adopted on the eight cylinders' surfaces. For the conformation tensor, the no-flux condition is approximated on all boundaries except for the inlet boundary with $\mathbf{c} = \mathbf{I}$.

The mesh is generated by the commercial software ICEM. As shown in Fig. 29, the surrounding region of each cylinder is discretized by the O-type mesh. The rest of the computational domain is discretized by several blocks of quadrilateral meshes, with the dense mesh near the cylinder region and the course mesh near the domain boundaries. Each block of the O-type mesh consists of 250 grid points uniformly distributed along the cylinder perimeter and 51 grid points stretched with an exponential progression along the radial direction to ensure a fine mesh near the cylinder surface. In this study, the size of the first cell adjacent to the cylinder surface in the radial direction is set to $0.0005D$. In the $x$ direction, 501 grid points (for $L_v$) are unevenly arranged in the downstream region and 71 grid points (for $L_u$) are set in the upstream region. The total mesh number is 244, 300. The Reynolds number is defined as $Re = U_\infty D \rho / \mu$ and fixed at $Re = 100$. The Weissenberg number is defined as $We = \lambda U_\infty / D$, which ranges from 0 to 10. To minimize the effect of the artificial dissipation, the artificial dissipation is set to zero for the cases with $We \leq 0.5$ and nonzero for the cases with $We > 0.5$.

The total lift coefficient ($C_l$) for the eight side-by-side cylinders is calculated as,

$$C_l = \frac{F_y}{4\rho U_\infty^2 D},$$ (10)

where $F_y$ is the total force acting on all the cylinders in the $y$ direction. The drag coefficient ($C_x$) and lift coefficient ($C_y$) for each cylinder are calculated as,

$$C_x = \frac{2F_x}{\rho U_\infty^2 D}, \quad \text{and} \quad C_y = \frac{2F_y}{\rho U_\infty^2 D},$$ (11)

where $F_x$ and $F_y$ are the hydrodynamic force acting on each cylinder in the $x$ and $y$ directions, respectively. The hydrodynamic force acting on a cylinder can also be decomposed into two components $F_\alpha$ and $F_n$, as shown in Fig. 30. $F_\alpha$ is the force component along the center-to-center line of the cylinders, which can be calculated as

$$F_\alpha = -F_x \cos\left(\frac{20^\text{o}}{180^\text{o}}\pi\right) + F_y \sin\left(\frac{20^\text{o}}{180^\text{o}}\pi\right).$$ (12)

The corresponding inter-cylinder interaction coefficient ($C_\alpha = 2F_\alpha \big/ \rho U_\infty^2 D$) could reflect the inter-cylinder interaction resulting from the hydrodynamic force.

The time-averaged lift coefficient ($\overline{C_l}$) of the eight cylinders in Newtonian and viscoelastic fluids are plotted in Fig. 31. In Newtonian fluid, $\overline{C_l}$ is 1.099. In viscoelastic fluid, $\overline{C_l}$ gradually increases with an increase in $We$. If the eight cylinders can be regarded as a single blunt body, the magnitude of the total lift coefficient could reflect the degree of the asymmetric distribution of hydrodynamic forces along this blunt body. The decreasing $\overline{C_l}$ indicates that the up and down asymmetry of the flow is weakened.

Moreover, the inter-cylinder interaction coefficient $\overline{C_\alpha}$ for each cylinder at different $We$ is listed in Tab. 3. $|\overline{C_\alpha}|$ of each cylinder always decreases with increasing $We$ except for cy-8. For example, $|\overline{C_\alpha}|$ of cy-1 is 6.499 in Newtonian fluid, while 3.754 for viscoelastic fluid of $We = 10$. The reduction in $|\overline{C_\alpha}|$ means that the inter-cylinder interaction is weakened, which is similar to viscoelastic flow over two side-by-side circular cylinders reported in our previous study.[33]

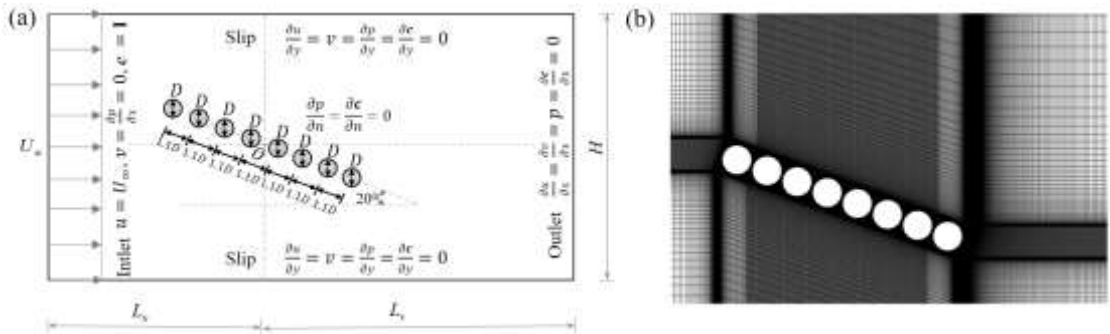

**Fig. 29.** (a) The schematic of unconfined flow over an inclined row of eight equally closely spaced circular cylinders. (b) The mesh topology near the cylinders.

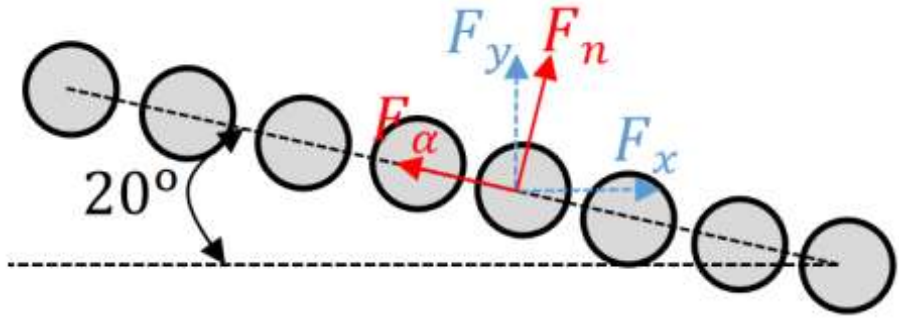

**Fig. 30.** Schematic of the hydrodynamic force acting on a cylinder. $F_x$ and $F_y$ denote the force components along the $x$ and $y$ directions, respectively. $F_\alpha$ and $F_n$ denote the force components along and perpendicular to the center-to-center line of the cylinders, respectively.

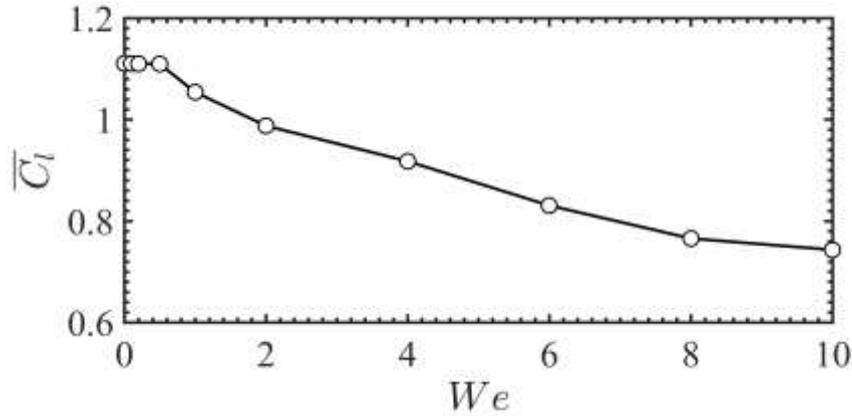

**Fig. 31.** Variation of the time-averaged lift coefficient with $We$.

**Table 3.** $C_\alpha$ of each cylinder at different $We$.

| Cylinders | cy-1 | cy-2 | cy-3 | cy-4 | cy-5 | cy-6 | cy-7 | cy-8 |
|-----------|------|------|------|------|------|------|------|------|
| Newtonian | -6.499 | -1.220 | -1.073 | -0.950 | -0.818 | -0.652 | -0.419 | 0.288 |
| $We = 0.1$ | -6.481 | -1.219 | -1.072 | -0.949 | -0.817 | -0.651 | -0.418 | 0.287 |
| $We = 0.2$ | -6.321 | -1.218 | -1.068 | -0.947 | -0.814 | -0.647 | -0.415 | 0.283 |
| $We = 0.5$ | -6.272 | -1.216 | -1.064 | -0.942 | -0.809 | -0.642 | -0.408 | 0.274 |
| $We = 1$ | -6.162 | -1.212 | -1.057 | -0.934 | -0.799 | -0.629 | -0.393 | 0.260 |
| $We = 2$ | -5.638 | -1.123 | -0.999 | -0.881 | -0.751 | -0.590 | -0.375 | 0.178 |
| $We = 4$ | -4.984 | -0.971 | -0.871 | -0.749 | -0.614 | -0.463 | -0.357 | -0.113 |
| $We = 6$ | -4.329 | -0.813 | -0.777 | -0.665 | -0.546 | -0.417 | -0.368 | -0.192 |
| $We = 8$ | -3.943 | -0.735 | -0.668 | -0.603 | -0.497 | -0.378 | -0.315 | -0.247 |
| $We = 10$ | -3.754 | -0.678 | -0.630 | -0.571 | -0.474 | -0.374 | -0.294 | -0.269 |

**Note:** Cylinders are numbered 1 to 8 from left to right. The negative sign before $C_\alpha$ denotes the leftward direction along the cylinder row.

Wake asymmetry weakening in viscoelastic fluids

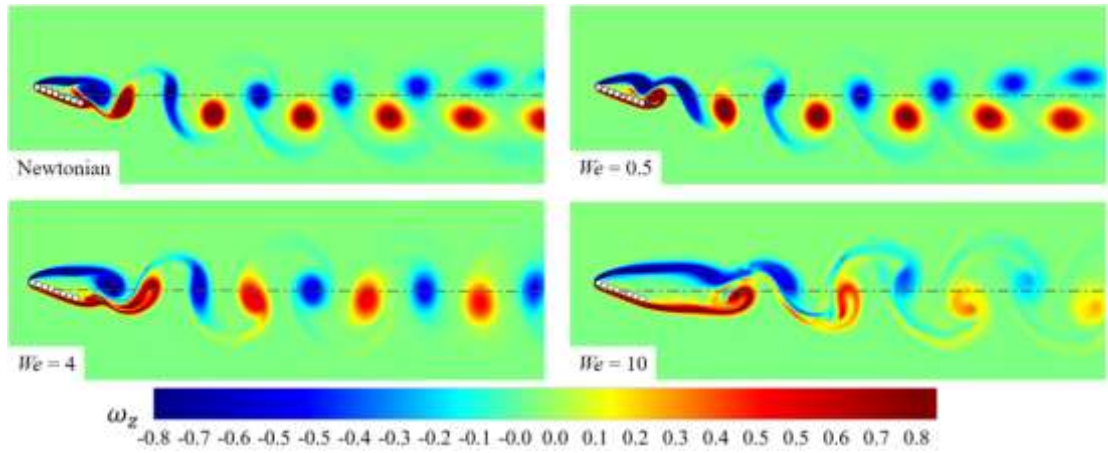

**Fig. 32.** The instantaneous vorticity contours for different *We*. $\omega_z$ is normalized by $U_\infty/D$.

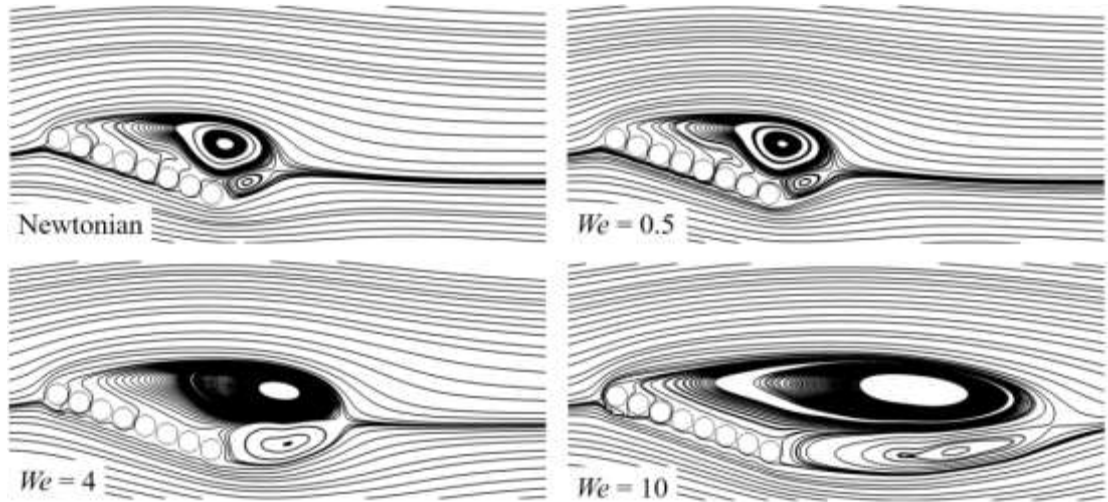

**Fig. 33.** The time-averaged streamlines for different *We*.

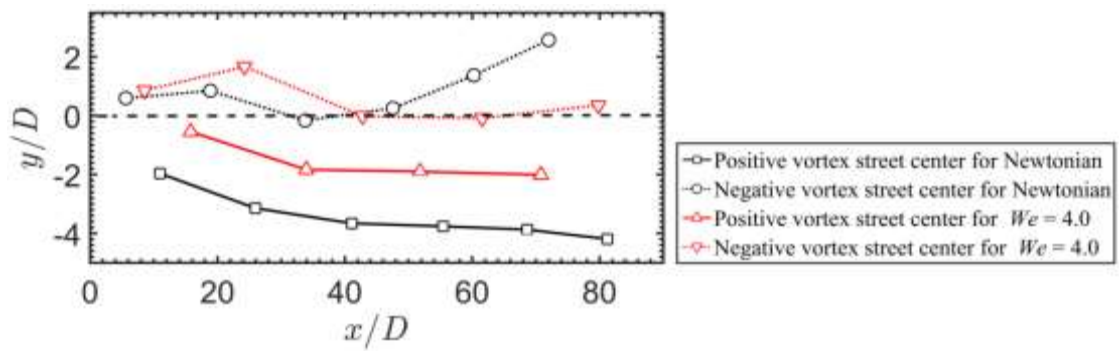

**Fig. 34.** The corresponding instantaneous positions of vortex cores for Newtonian and viscoelastic flows shown in Fig. 32.

Wake asymmetry weakening in viscoelastic fluids

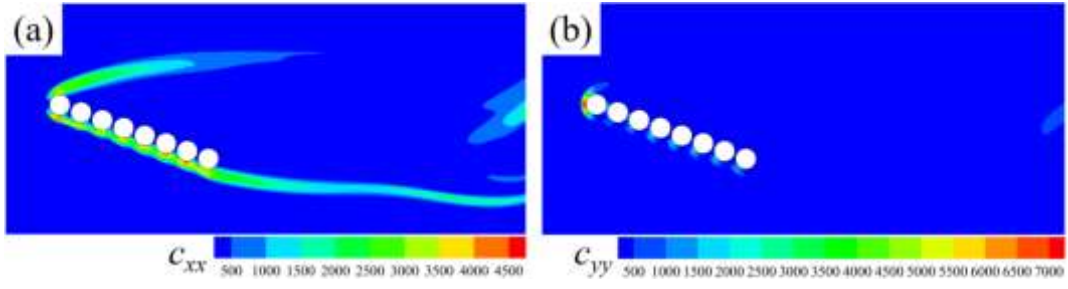

**Fig. 35.** The instantaneous distributions of (a) $c_{xx}$ and (b) $c_{yy}$ at $We = 10$.

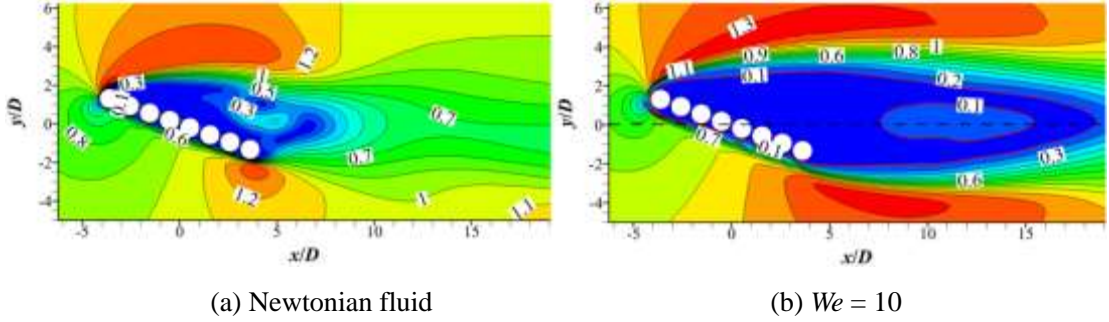

(a) Newtonian fluid                    (b) $We = 10$

**Fig. 36.** The distribution of time-averaged flow velocity magnitude ($\sqrt{\overline{u}^2 + \overline{v}^2}/U_\infty$) for (a) Newtonian fluid and (b) viscoelastic fluid with $We = 10$ and $L = 100$. The contour of $\sqrt{\overline{u}^2 + \overline{v}^2}/U_\infty = 0.1$ is marked with the pink line in panel (b).

The instantaneous vorticity contours and the time-averaged streamlines are shown in Fig. 32 and Fig. 33, respectively. The corresponding instantaneous positions of vortex cores are extracted and plotted in Fig. 34. For both Newtonian and viscoelastic fluids, the flow resembles flow over a single blunt body. In Newtonian fluid, the vortex shedding trajectory deviates from the horizontal center line of the computational domain (the dash-dotted line in Fig. 32). In viscoelastic fluid, the vortex shedding trajectory gradually approaches the center line as $We$ increases. At the same time, the time-averaged recirculation wake is elongated and tends to be symmetric against the center line.

For viscoelastic flow over an inclined row of eight equally closely spaced circular cylinders, the elastic stress is higher near the front edge of each cylinder. The regions with higher elastic stress connect into one piece (similar to viscoelastic flow over side-by-side circular cylinders when $L_D = 0.1D$[33]), which further extends downstream along the lower side of each cylinder, as shown in Fig. 35. When encountering the elastic boundary, the flow becomes slower. The corresponding distribution of the time-averaged flow velocity magnitude ($\sqrt{\overline{u}^2 + \overline{v}^2}/U_\infty$) is shown in Fig. 33(b), which is compared with that of Newtonian fluid, as shown in Fig. 36(a). The contour of $\sqrt{\overline{u}^2 + \overline{v}^2}/U_\infty = 0.1$ is marked with the pink line Fig. 36(b). The flow velocity within the region enclosed by the pink line is lower than 0.1, which could be regarded as a whole blunt body. The whole blunt body is symmetric along the center line of the flow field, which results in a symmetric vortex shedding trajectory along the flow field center line. Specially, the time-averaged velocity magnitude is uniform and nearly zero in the vicinity of each cylinder. The shear stress and pressure distributes symmetrically for the lower and upper surfaces of each cylinder, which results in a low inter-cylinder interaction.

Wake asymmetry weakening in viscoelastic fluids

### E. Common feature for these four serial cases

A common feature can be summarized from the above stress distributions for the four examples, i.e., a region with high elastic stress similar to shock wave appears near the frontal edge of the blunt body and extends downstream, as sketched by the blue solid lines in Fig. 37. This region acts as a stationary shield to divide the flow into different regimes. Thus, the free stream resembles passing this shield instead of the original blunt body. As this region is nearly symmetric against the horizontal center line of the corresponding flow configuration, the wake flow recovers symmetry. In particular, another high elastic stress region surrounds the rotating circular, which results in the uniform velocity along the circumferential direction within the rotating boundary layer. The whole rotating boundary layer behaves like a larger non-rotating cylinder (the dotted line in Fig. 37(b)), resulting in a symmetric wake field downstream.

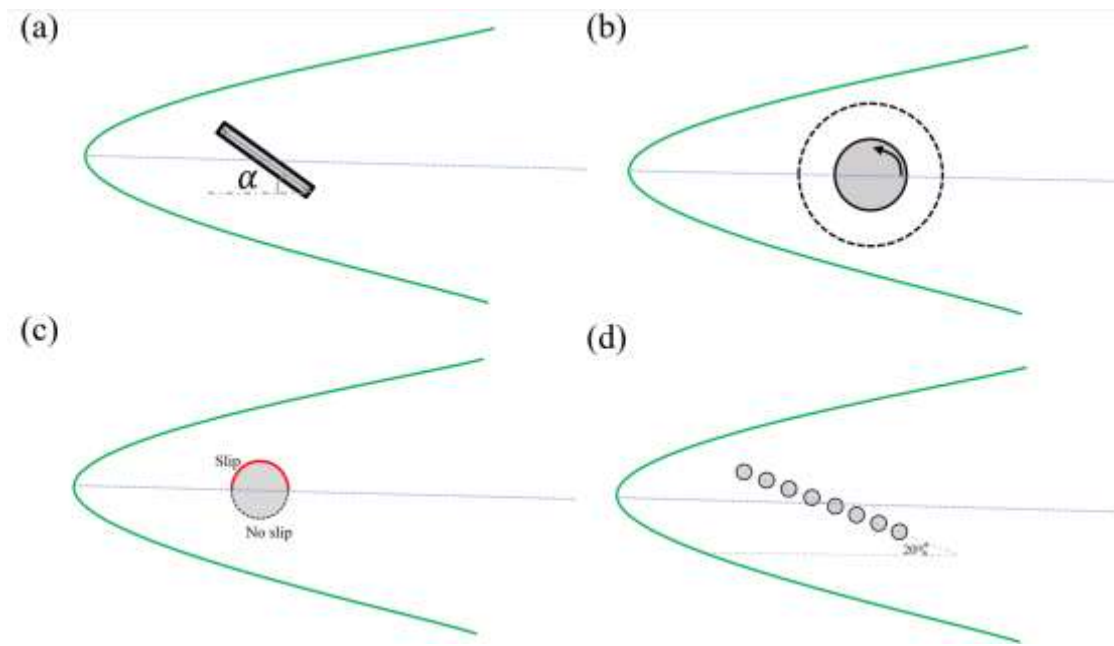

**Fig. 37.** Schematic diagram of regions with high elastic stress (the blue solid lines): (a) flow over an inclined flat plate with various angles of incidence, (b) flow over a rotating circular cylinder, (c) flow over a circular cylinder with asymmetric slip boundary distribution, and (d) flow over an inclined row of eight equally closely spaced circular cylinders.

### V. Conclusion Remarks

In this paper, we report that viscoelastic wake flow asymmetry in a free domain is no longer sensitive to asymmetry in geometry or boundary condition through numerical simulation. Two-dimensional direct numerical simulations based on the finite-extensible nonlinear elastic model with the Peterlin closure (FENE-P model) are conducted. Four asymmetric flows are considered in this study, i.e., flow over an inclined flat plate with various angles of incidence at $Re$ = 500, flow over an inclined row of eight equally closely spaced circular cylinders at $Re$ = 100, flow over a rotating circular cylinder at $Re$ = 100 and flow over a circular cylinder with asymmetric slip boundary distribution at $Re$ = 500. The four flows are the representatives of geometric asymmetry and boundary condition asymmetry.

The lift coefficient can be used to describe the asymmetry degree of the flow field near the blunt body. The lift coefficient acted on blunt body in viscoelastic fluid becomes lower than that in Newtonian fluid. Accordingly, the viscoelastic wake flow tends to be symmetric with respect to the horizontal center line of the flow configuration for inclined flat plate with an angle of incidence, the rotating circular cylinder, a circular cylinder with asymmetric slip boundary distribution or an inclined row of eight equally closely spaced circular cylinders. Specially, for the multibody flow with closely placed objects, the time-averaged velocity magnitude is uniform and nearly zero in the vicinity of each object, which results in a low inter-object interaction.

A careful examination of the elastic stress distributions for the four examples reveals a common feature, i.e., an region with high elastic stress similar to shock wave appears near the frontal edge of the blunt body and extends all the way downstream. The blunt body seems to be wrapped by this region, which retains a symmetric shape with respect to the horizontal center line of the flow configuration. The free stream cannot directly interact with the blunt body but flows around this region, as if the free stream passes an imaginary symmetric blunt body with larger characteristic length. Thus, the wake flow recovers symmetry. In particular, another high elastic stress region surrounds the rotating circular, which results in a uniform tangential velocity at a specific radial location with respect to the center of the cylinder. This tangential velocity becomes zero at the outer of the rotating boundary layer. Thus, the external flow behaves like flow past a larger non-rotating cylinder and exhibits a symmetric wake field downstream.

The influence of elastic stress distribution on lift and flow symmetry discovered in this paper could be used for precise flow control. It would be interesting to investigate how to obtain the desired lift coefficient and vortex shedding trajectory by controlling the polymer concentration distribution[40] around an inclined flat plate or a hydrofoil to redistribute elastic stress in future. The flow asymmetry could cause higher flow-induced vibration response for a rotating circular cylinder.[61] Xiong et al.[15] proposed that polymer addition could restrain vortex-induced vibration of a non-rotating cylinder. It is reasonable to speculate that polymer addition could also effectively suppress vortex-induced vibration of a rotating circular cylinder due to the flow asymmetry weakening discussed above, which entails future investigations.

## Data Availability Statement

The data that support the findings of this study are available from the corresponding author upon reasonable request.

## Acknowledgement

The author Peng Yu would like to thank the financial support from the National Natural Science Foundation of China (NSFC, Grant Nos. 12172163, 12071367). The author Xiao-ru Zhuang would like to thank the Guangdong Basic and Applied Basic Research Foundation (Grant No. 2022A1515011057). This work is supported by Center for Computational Science and Engineering of Southern University of Science and Technology. The author Sai Peng would like to thank Dr. Jun-qiang Zhang from Huazhong University of Science and Technology for useful discussion.

Wake asymmetry weakening in viscoelastic fluids